\documentclass[aps,prc,twocolumn,showpacs,preprintnumbers,nofootinbib,float,superscriptaddress]{revtex4-1}

\usepackage{epsfig}
 
\usepackage[dvipsnames]{xcolor}
\usepackage{float,amsmath,amssymb}
\usepackage{graphicx}
\usepackage{url}
\usepackage{bbold}
\usepackage[utf8]{inputenc}
\usepackage{physics}
\newcommand{\der}{\mathrm{d}}

\newcommand{\xpom}{x_\mathbb{P}}

\newcommand{\rt}{{\mathbf{r}_\perp}}

\newcommand{\xt}{{\mathbf{x}_\perp}}
\newcommand{\kt}{{\mathbf{k}_\perp}}

\newcommand{\bt}{{\mathbf{b}_\perp}}
\newcommand{\bti}{{\mathbf{b}_{\perp,i}}}
\newcommand{\Deltat}{{\boldsymbol{\Delta}_\perp}}

\newcommand{\nc}{{N_\mathrm{c}}}
\newcommand{\jpsi}{$\mathrm{J}/\psi$ }
\newcommand{\jpsim}{\mathrm{J}/\psi}

\newcommand{\aem}{\alpha_\mathrm{em}}
\newcommand{\Bt}{\mathbf{B}_{\perp}}

\newcommand{\qt}{\mathbf{q}_{\perp}}
\newcommand{\Mcal}{\mathcal{M}}
\newcommand{\Fcal}{\mathcal{F}}
\newcommand{\Acal}{\mathcal{A}}
\newcommand{\Pt}{\mathbf{P}_{\perp}}

\newcommand {\RuRu}	{$^{96}_{44}$Ru+$^{96}_{44}$Ru}
\newcommand {\ZrZr}	{$^{96}_{40}$Zr+$^{96}_{40}$Zr}
\usepackage[breaklinks,colorlinks,citecolor=citcolor,urlcolor=blue,linkcolor=lcolor]{hyperref}
\definecolor{lcolor}{rgb}{0.5,0,0}
\definecolor{citcolor}{rgb}{0,0.3,0.0}

\begin{document}

\title{Effects of nuclear structure and quantum interference on diffractive vector meson production in ultra-peripheral nuclear collisions}

\author{Heikki M\"antysaari}
\affiliation{Department of Physics, University of Jyv\"askyl\"a, P.O. Box 35, 40014 University of Jyv\"askyl\"a, Finland}
\affiliation{Helsinki Institute of Physics, P.O. Box 64, 00014 University of Helsinki, Finland}

\author{Farid Salazar}
\affiliation{Nuclear Science Division, Lawrence Berkeley National Laboratory, Berkeley, California 94720, USA}
\affiliation{Physics Department, University of California, Berkeley, California 94720, USA}
\affiliation{Department of Physics and Astronomy, University of California, Los Angeles, California 90095, USA}
\affiliation{Mani L. Bhaumik Institute for Theoretical Physics, University of California, Los Angeles, California 90095, USA}

\author{Bj\"orn Schenke}
\affiliation{Physics Department, Brookhaven National Laboratory, Upton, NY 11973, USA}

\author{Chun Shen}
\affiliation{Department of Physics and Astronomy, Wayne State University, Detroit, Michigan 48201, USA}
\affiliation{RIKEN BNL Research Center, Brookhaven National Laboratory, Upton, NY 11973, USA}

\author{Wenbin Zhao}
\affiliation{Department of Physics and Astronomy, Wayne State University, Detroit, Michigan 48201, USA}
\affiliation{Nuclear Science Division, Lawrence Berkeley National Laboratory, Berkeley, California 94720, USA}
\affiliation{Physics Department, University of California, Berkeley, California 94720, USA}

\begin{abstract}
We study diffractive vector meson production in ultra-peripheral collisions (UPCs) of heavy nuclei, utilizing a theoretical framework based on the Color Glass Condensate (CGC) formalism. We focus on Au+Au, U+U, Ru+Ru,  Zr+Zr, and Pb+Pb collisions, examining the transverse momentum dependence of vector meson production cross-sections and ${\rm cos(2\Delta\Phi)}$ asymmetries in the decay product distributions to explore the role of nuclear geometry. The angular modulation is due to the linear polarization of the incoming photons and quantum interference effects. We extract nuclear radii and find them to be consistent with experimental data from the STAR collaboration. The amplitudes of the ${\rm cos(2\Delta\Phi)}$ modulation in the cross-section and the extracted radii depend on the nuclear geometry. This dependence is dominated by the geometry-dependent variation of the minimum impact parameter required for ultra-peripheral collisions.
\end{abstract}
\maketitle


\section{Introduction}

Exploring nuclear deformation has been a central topic in nuclear physics. The understanding of nuclear structure is essential as it plays a fundamental role in a wide range of phenomena, from nuclear reactions and element synthesis in stars to the behavior of matter in extreme astrophysical environments \cite{osti_1999724}.
Recently, new methods for accessing nuclear structure in high-energy collision experiments have emerged, including heavy ion collisions \cite{Giacalone:2019pca,Giacalone:2021uhj,Bally:2022vgo} and future electron-ion collisions \cite{Mantysaari:2023qsq}.

Heavy-ion collisions are used to explore nuclear matter under extreme conditions and have been shown to create the deconfined Quark-Gluon Plasma (QGP) \cite{Harris:1996zx,Harris:2023tti}. In these collisions, the initial shape of the colliding nuclei plays a crucial role in determining the spatial distribution of nuclear matter produced in the collision. This distribution determines the initial pressure anisotropies, which are transformed into observable momentum space correlations via strong final state interactions. Hydrodynamic modeling of heavy-ion collisions, incorporating a precise description of the initial state of different ion species, has demonstrated that the detailed structural properties of the colliding nuclei affect a variety of observables \cite{Filip:2009zz,Masui:2009qk,Hirano:2012kj,Shen:2014vra,Schenke:2014tga,Mantysaari:2017cni,Schenke:2020mbo,Xu:2021uar,Giacalone:2021udy,Zhang:2021kxj,Nijs:2021kvn,Zhao:2022mce,Ryssens:2023fkv}. 

Future Deep Inelastic Scattering (DIS) facilities, such as the Electron-Ion Collider (EIC)~\cite{AbdulKhalek:2021gbh,Aschenauer:2017jsk}, LHeC/FCC-he~\cite{LHeC:2020van}, and EicC~\cite{Anderle:2021wcy}, hold great promise in their pursuit to unravel the multi-dimensional structure of protons and nuclei. In particular, it has been recently demonstrated that exclusive vector meson production at high energy in electron-nucleus ($e+A$) collisions can provide valuable insights into the geometric structure of the target nucleus at multiple length scales, including its deformation~\cite{Mantysaari:2023qsq}. One advantage of $e+A$ over heavy-ion collisions is that it provides a cleaner probe of the nuclear properties.

Before these facilities are realized, ultra-peripheral collisions (UPCs) of heavy ions at the Relativistic Heavy Ion Collider (RHIC) and the Large Hadron Collider (LHC) offer a great opportunity to explore nuclear structure with beams of quasi-real photons~\cite{Bertulani:2005ru,Klein:2019qfb}. In UPCs, for which the distance between the two colliding nuclei is larger than the sum of the radii of the nuclei,\footnote{This statement is approximate as, in any given event, there can be fluctuations of which impact parameters allow for a UPC, as we will see below.} the strong hadronic interactions are suppressed and the photon-nucleus ($\gamma+A$) interactions involving photons emitted from one of the colliding nuclei are dominant. UPCs provide high-energy quasi-real photons at a high luminosity, allowing for the study of exclusive vector meson production with high precision. 

In the dipole picture, valid at high energy, the production of vector meson in UPCs occurs as follows: the quasi-real photon splits into a quark-antiquark pair, which then interacts with the nuclear target via a net color-neutral exchange. The transverse momentum $\Deltat$ imparted by the nuclear target is Fourier conjugate to the impact parameter of the photon relative to the center of the target nucleus, allowing the spatial imaging of gluons inside the nuclei. Experimentally, one measures (reconstructs) the transverse momentum $\qt$ of the vector meson, which by momentum conservation is the sum of $\Deltat$ and $\kt$, the latter being the transverse momentum carried by the quasi-real photon. 
The inability to distinguish which nucleus is the photon source or the target, results in the quantum interference of the two contributions
\cite{Bertulani:2005ru}. A proper description of the experimental measurement of vector meson production and decay requires the consideration of the photon $\kt$, the quantum interference effects, as well as the linearly polarized nature of the incoming photons involved \cite{Xing:2020hwh,Brandenburg:2022jgr}.

Recent observations by the STAR collaboration have revealed significant ${\rm cos} (2\Delta\Phi)$ modulations in exclusive $\rho$ meson photoproduction in UPCs~\cite{STAR:2022wfe}, where $\Delta\Phi$ is the angle between the produced $\rho$ meson transverse momentum $\mathbf{q}_\perp$ and its decay product pions' relative transverse momentum. 
The observed modulations arise due to the linear polarization of the UPC photons, as well as the interference between the two indistinguishable channels.

STAR has used this data to extract the strong-interaction nuclear radii of ${\rm ^{197} Au }$ and ${\rm ^{238} U }$. For both nuclei, radii larger than the charge radius were found. For uranium, the obtained values were even larger than previous results that take into account the presence of a neutron skin. Recently, STAR presented a new measurement of the $\cos(2\Delta\Phi)$ modulation for diffractive \jpsi production in ultra-peripheral Au+Au collisions at Quark Matter 2023~\cite{STAR:2023ashik}. In addition, STAR has reported results for vector meson production in ultra-peripheral isobar \RuRu\ and \ZrZr\ collisions \cite{STAR:2023zhao}, which show differences attributed to the different radii and deformations of Ru and Zr. The isobar collisions were originally proposed to control the background in search for the chiral magnetic effect (CME)~\cite{Kharzeev:2007jp,Voloshin:2010ut,Kharzeev:2015znc,Deng:2016knn,Zhao:2019hta,STAR:2019bjg,STAR:2021mii}. However, the nuclear structure difference caused significant observable differences between the isobar systems, such as in their event multiplicities and elliptic flow coefficients~\cite{Xu:2017zcn,Nijs:2021kvn}. As these differences were not considered before the blind analysis, the original background estimates were not sufficient to allow for a CME signal extraction. This highlights the importance of a detailed understanding of nuclear structure.

In this work, we investigate the $q^2_{\perp}$-dependence of the diffractive vector meson production cross-section and the ${\rm cos(2\Delta\Phi)}$ modulation in the angular distributions of decay products in ultra-peripheral Au+Au, U+U, Zr+Zr, Ru+Ru, and Pb+Pb collisions within the Color Glass Condensate (CGC) framework.  
We demonstrate how the deformation of the uranium nucleus impacts the $q^2_{\perp}$-dependence of the cross-section and the extracted radius for $\rho$ production. We further show predictions for $\rho$, $\phi$ and \jpsi production and their ${\rm cos(2\Delta\Phi)}$ modulation in ultra-peripheral Au+Au collisions.  We discuss the sensitivity of the $q^2_{\perp}$-dependent ${\rm cos(2\Delta\Phi)}$ modulation and extracted radius to the minimum impact parameter $B_{\rm min}$.  We also study the ${\rm cos(2\Delta\Phi)}$ modulations in different forward neutron multiplicity classes in Pb+Pb collisions at the LHC and compare them to experimental data from the ALICE Collaboration.
Finally, we present the sensitivity of the production of vector mesons in ultra-peripheral isobar collisions to the structure of Ru and Zr nuclei.

This paper is organized as follows. In Sec.~\ref{sec:upc} we review the formalism for computing exclusive vector meson production cross-sections in ultra-peripheral collisions, including the decay of the vector meson using a joint impact parameter and transverse momentum-dependent framework. The CGC implementation of the calculation of the scattering amplitudes for vector meson production is presented in Sec.~\ref{sec:vm_production}. 
Our results are presented, compared with data (where available), and discussed in Sec.~\ref{sec:results}. We conclude by presenting a summary of our findings in Sec.~\ref{sec:conclusions}.
\\~\\

\section{Ultra peripheral collisions}
\label{sec:upc}

As mentioned in the introduction, in UPCs there are two indistinguishable contributions to vector meson production. This is because either of the colliding nuclei can act as a source of linearly polarized photons while the other plays the role of the target. As a consequence, we need to take into account the quantum mechanical interference of these two processes~\cite{Bertulani:2005ru}.
Another important ingredient is the implementation of the non-zero transverse momentum of the incoming photons, as it significantly impacts the transverse momentum spectrum of the produced vector mesons. This effect is particularly important near the diffractive minima of the coherent cross-section.

Following the joint impact parameter and transverse momentum-dependent framework in Ref.~\cite{Xing:2020hwh,Hagiwara:2020juc}, 
the differential cross-section for the diffractive $\rho$ or $\phi$ production, and the subsequent decay into pions or kaons, respectively, is given by
\begin{widetext}
\begin{align}
\label{eq:crossection}
    &\frac{\der \sigma^{ \rho \to \pi^+ \pi^- (\phi\to K^+K^-)}}{\der^2 \Pt \der^2 \qt \der y_1 \der y_2 }  = \nonumber \\
    & \frac{1}{2 (2\pi)^3} \frac{f^2}{(Q^2 -M_V^2)^2 + M_V^2 \Gamma^2}  \Bigg \{\left<  \int \der^2 \Bt  \Mcal^{i}(y_1, y_2,\qt,\Bt)\Mcal^{\dagger,j}(y_1, y_2,\qt,\Bt)  \Pt^i \Pt^j  \Theta(|\Bt|-B_{\rm min})\right>_{\Omega}\Bigg \} \,. 
\end{align}

\end{widetext}
This expression takes into account the (four-)momentum transfer $k$ imparted by the linearly polarized quasi-real photons and the momentum transfer $\Delta$ of the nuclear target, both of which contribute to the momentum of the produced vector meson $q=k+\Delta$. The information on these momenta is encoded in the amplitude $\mathcal{M}^i$ (see below). This framework also implements the interference effect in the exchange of photon source and target. The transition $\rho\rightarrow\pi^++\pi^-$, or $\phi\to K^++K^-$ follows the usual Breit-Wigner approach, implemented at the amplitude level \cite{Hagiwara:2020juc}.

Here $M_V$ denotes the vector meson mass, $Q$ is the invariant mass of the daughter particle system, and $y_1$ and $y_2$ are the daughter particles’ rapidities. $\Pt =(\textbf{p}_{1\perp}-\textbf{p}_{2\perp})/2$ and $\qt = \textbf{p}_{1\perp} + \textbf{p}_{2\perp}$ with $\textbf{p}_{1\perp}$ and $\textbf{p}_{2\perp}$ being the transverse momenta of the daughter particles. An important consequence of the linear polarization of the photons is its imprint on the angular correlations between the transverse momenta of the daughter particles $\Pt$ and $\qt$, which in Eq.\,\eqref{eq:crossection} are reflected in the tensor contraction between $ \Mcal^{i}\Mcal^{\dagger,j}$ and $\Pt^i \Pt^j$ \cite{Xing:2020hwh,Mantysaari:2022sux}.

\begin{figure}
    \centering
    \includegraphics[width=0.47\textwidth]{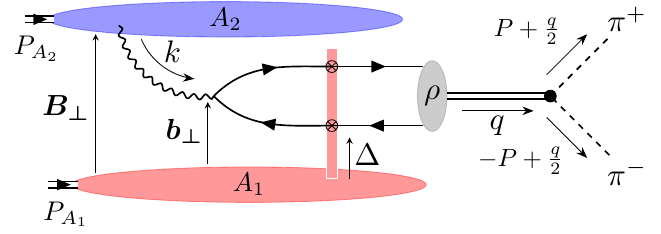}
    \caption{Schematic of $\rho$ meson production in UPCs. The red rectangle represents the CGC shock-wave interaction of the quark-antiquark pair with the target. The gray oval represents the formation of the $\rho$ meson, and the black dot represents its decay into a $\pi^+ \pi^-$ pair. We only show the diagram in which the nucleus $A_1$ acts as the target, while the nucleus $A_2$ acts as photon source. $\Bt$ and $\bt$ are transverse position vectors of the center of the photon source (ion $A_2$) and the photon, respectively, relative to the center of the target (ion $A_1$).
    }
\label{fig:schematic_vector_meson_production}
\end{figure}

We use $\Gamma=\Gamma_{\rho}=0.156 \ {\text {GeV}}$ \cite{Flores-Baez:2007vnd,Hagiwara:2020juc,ParticleDataGroup:2018ovx}, and $\Gamma_{\phi}= 0.00443 \,{\text {GeV}}$~\cite{Flores-Baez:2007vnd, ParticleDataGroup:2014cgo} for $\rho$ and $\phi$ production, respectively. The effective couplings $f$ are not used as the available experimental data does not determine the absolute normalization. The minimum impact parameter above which hadronic collisions seize to occur is denoted by $ B_{\rm min}$. It is determined by requiring no nucleon-nucleon collision (${N_{\rm coll}} = 0$) event-by-event, employing the Monte Carlo Glauber model.
The average over the target configurations $\Omega$ is taken at the cross-section level, which corresponds to the total diffractive (sum of coherent and incoherent) cross-section. Consequently, our event-by-event calculation includes the interference term for both coherent and incoherent contributions. The interference term has a negligible contribution in the $q_\perp$ region where the incoherent cross-section dominates over the coherent part. Whether or not the interference is included in the incoherent cross-section results in a small difference.

It is advantageous for our subsequent numerical implementation to express the amplitude $\Mcal^{i}(y_1, y_2,\qt,\Bt)$ as the convolution of the photon field $\widetilde{\Fcal}_{A_k}^i(x_k,\bt-\Bt)$ and the vector meson production amplitude $\widetilde{\Acal}_{A_k}(x_k,\bt)$ in coordinate space \cite{Mantysaari:2022sux}:
\begin{widetext}
\begin{align}
     \Mcal^i(y_1,y_2,\qt,\Bt) 
     = \int \der^2 \bt e^{-i\qt\cdot\bt} \left[ \widetilde{\Acal}_{A_1}(x_1,\bt) \widetilde{\Fcal}_{A_2}^i(x_2,\bt-\Bt) + 
     \widetilde{\Acal}_{A_2}(x_2,\bt-\Bt) \widetilde{\Fcal}_{A_1}^i(x_1,\bt) \right].
     \label{eq:amplitude_coordinate_space}
\end{align}
\end{widetext}
The subscripts $A_1$ and $A_2$ refer to the colliding nuclei, and the $x$ values are  $x_1 = \sqrt{(P_\perp^2 + m^2)/s} (e^{y_1} + e^{y_2}) $ and $x_2 = \sqrt{(P_\perp^2 + m^2)/s} (e^{-y_1} + e^{-y_2}) $, where $m$ is the mass of the daughter particle.
The first term in the square bracket of Eq.\,\eqref{eq:amplitude_coordinate_space}, corresponds to the contribution in which $A_2$ is the photon source and $A_1$ is the nuclear target, while the second term corresponds to the inverse. $\Bt$ is the impact parameter vector between the two nuclei, and $\bt$ is the impact parameter vector of the quasi-real photon relative to the center of the target nucleus. The functions $\widetilde{\Fcal}^j_{A_k}$ and $\widetilde{\Acal}_{A_k}$ are the Fourier transforms of the electromagnetic field $\Fcal^j_{A_k}(x_k,\kt)$ of the photon source and the amplitude for vector meson production $\Acal^j_{A_k}(x_k,\Deltat)$, respectively. The latter is calculated in the CGC formalism.

We will specify the amplitude $\widetilde{\Acal}^j_{A_k}(x_k,\bt)$ in detail in Sec.\,\ref{sec:vm_production}. As we only need the electromagnetic field at distances $\vert\Bt\vert \gtrsim B_{\rm min} $, we can replace the electromagnetic field of the nucleus as that of a point particle of charge $Z_A$ moving relativistically (using a Woods-Saxon form factor has a negligible effect on our results). In this case, we have
\begin{align}
    \widetilde{\Fcal}_A^j(x_k,\bt)  =\frac{1}{2\pi} \frac{Z_A \alpha_\mathrm{em}^{1/2} }{\pi} \frac{\bt^j}{|\bt|} x_k M_p K_1\left( x_k M_p |\bt|\right) \,,
    \label{eq:EM_fieldZ_Gauss}
\end{align}
where $Z_A$ is the ion charge, and $M_p$ is the mass of the nucleon.

In order to make the $\cos (2\Delta\Phi)$ modulation more clearly visible, we 
introduce the isotropic ($C_0$) and elliptic ($C_2$) coefficients and write Eq.~\eqref{eq:crossection} as 
\begin{widetext}
\begin{align}
    \frac{\der \sigma^{ \rho \to \pi^+ \pi^- (\phi\to K^+ K^-)}}{\der^2 \Pt \der^2 \qt \der y_1 \der y_2 }  
     = \frac{1}{4 (2\pi)^3} \frac{P_\perp^2 f^2}{(Q^2 -M_V^2)^2 + M_V^2 \Gamma^2}  \Bigg \{ C_0(x_1,x_2,|\qt|)  +  C_2(x_1,x_2,|\qt|) \cos(2 (\phi_{\Pt} -\phi_{\qt})) \Bigg \} \,,
\end{align}
where
\begin{align}
 C_0(x_1,x_2,|\qt|) &=  \left<\int \der^2 \Bt \Mcal^i(x_1,x_2,\qt,\Bt) \Mcal^{\dagger,i} (x_1,x_2,\qt,\Bt) \Theta(|\Bt| - B_{\rm min})\right>_{\Omega}  \label{eq:C0} \,,\ \text{and} \\
 C_2(x_1,x_2,|\qt|) &= \left( \frac{2 \qt^i \qt^j}{\qt^2} - \delta^{ij} \right) \left<\int \der^2 \Bt \Mcal^i(x_1,x_2,\qt,\Bt) \Mcal^{\dagger,j} (x_1,x_2,\qt,\Bt) \Theta(|\Bt| - B_{\rm min}) \right>_{\Omega}\label{eq:C2} \,.
\end{align}
Similarly, the diffractive \jpsi production with subsequent decay into a dilepton can be written as 
\begin{align}
    &\frac{\der \sigma^{ \jpsim \to l^+ l^- }}{\der^2 \Pt \der^2 \qt \der y_1 \der y_2 }  \nonumber \\ 
    &= \frac{24 \alpha_{\mathrm{em}}^2 e_q^2}{(Q^2 -M_V^2)^2 + M_V^2 \Gamma^2} \frac{|\phi_{\jpsim}(0)|^2}{\pi M_V}  \Bigg \{ \left[ 1 - \frac{2 \Pt^2}{M_V^2} \right] C_0(x_1,x_2,|\qt|)  - \frac{2 \Pt^2}{M_V^2} C_2(x_1,x_2,|\qt|) \cos(2 (\phi_{\Pt} -\phi_{\qt})) \Bigg \}\,,
    \label{eq:Jpsi-cross-section}
\end{align}
\end{widetext}
where $|\phi_{\jpsim}(0)| = 0.0447~{\rm GeV^3}$ \cite{ParticleDataGroup:2018ovx} is the value of the modulus squared of the radial wave function of the \jpsi at the origin.  The \jpsi EM decay width into two leptons is related to the zero point wave function through $\Gamma=16 \pi \aem^2 e_q^2 \frac{|\phi_{\jpsim}(0)|^2}{M_V^2} $ \cite{ParticleDataGroup:2018ovx}, with $e_q= 2/3$ the charm quark charge in units of $e$, and $\aem = 1/137$ the fine-structure constant. Eq.\,\eqref{eq:Jpsi-cross-section} is equivalent to the result derived in \cite{Brandenburg:2022jgr}.

The difference in the angular dependence $\Delta\Phi$ of the differential cross-section for \jpsi and $\rho$ or $\phi$  production can be attributed to the different spin of the decay products. The \jpsi decays into electrons or muons, which are spin 1/2 particles, while $\rho$ and $\phi$ decay into pions or kaons, which are scalar particles \cite{Hagiwara:2020juc, Brandenburg:2022jgr}.  

Note that with the definitions of $C_0$ and $C_2$, the ${\rm cos(2\Delta\Phi)}$ asymmetry can be expressed as: 
\begin{align}\label{eq:modulation}
    \langle \cos[2(\phi_{\Pt}-\phi_{\qt})]\rangle =  \frac{1}{2} \frac{\int {\cal V_{\rm 2}}|\Pt|\dd |\Pt| C_2(|\qt|)}{\int {\cal V_{\rm 0}}|\Pt| \dd|\Pt|C_0(|\qt|)} ,
\end{align}
where ${\cal V_{\rm 2}} = {\cal V_{\rm 0}}= 1$ for $\rho\to\pi^+\pi^-$ or $\phi\to K^+K^-$, and ${\cal V_{\rm 2}} = -2\Pt^2$, ${\cal V_{\rm 0}} = (M_V^2-2\Pt^2)$ for $\jpsim \to l^+l^-$. Here the $x$ dependence of $C_0$ and $C_2$ in Eqs. \eqref{eq:C0} and \eqref{eq:C2} is removed by integrating  $y_1$ and $y_2$ over the experimental kinematic range.

In the case of $\rho$ and $\phi$ there are potential problems with contributions from non-perturbative physics that are not well under control.  
Due to the relatively small vector meson masses, there is no large scale ensuring that only small dipoles in the perturbative domain contribute. 
This could be remedied at high energy, where saturation scales are large and cross-sections will be dominated by contributions from small dipoles.
We will study the sensitivity of our results in realistic kinematics to the large-dipole contributions below.

Classification of UPC events is possible by detecting neutrons emitted at forward angles.  The probability for the target nucleus to emit a neutron is strongly dependent on the impact parameter. This leads to a larger weight for smaller impact parameters when more neutrons are detected. To account for this effect, we integrate the differential cross-section over an impact parameter range from $B_{\rm min}$ to $\infty$ with a weight function, 
\begin{eqnarray}
2\pi\int_{B_{\rm min}}^{\infty} B_\perp \der B_\perp P^2(B_\perp) \der \sigma(B_\perp, ...) \,,
\end{eqnarray} 
where the probability $P(B_\perp)$ of emitting a neutron from a nucleus is conventionally parametrized as~\cite{Baur:1998ay,Xing:2020hwh}
$P(B_\perp)= P_{1n}(B_\perp) \exp \left [-P_{1n}(B_\perp)\right ]$
which is denoted as the ``1n'' event, while for emitting any number of neutrons (``Xn'' event),  the probability is given by $P(B_\perp)= 1-\exp \left [-P_{1n}(B_\perp) \right ]$
with\footnote{There is some uncertainty related to the prefactor in Eq.~\eqref{eq:neutronemission} reflected by different numerical values quoted in Refs.~\cite{Baltz:1997di} and \cite{Xing:2020hwh}. We further note that neither of these values accurately reproduces the photon fluxes quoted in recent ALICE papers~\cite{ALICE:2023jgu,ALICE:2023gcs} obtained from the {\bf $\mathrm{n_O^On}$} (noon)~\cite{Broz:2019kpl} Monte Carlo setup. } 
\begin{equation} \label{eq:neutronemission}
P_{1n}(B_\perp)= 5.45\times10^{-5} \frac{Z^3(A-Z) }{A^{2/3} B_\perp^2 } \ \text{fm}^2.\end{equation} 
To compare to STAR measurements, the subsequent numerical calculations for Au+Au and U+U \cite{STAR:2022wfe} are made for ``XnXn'' events. For isobar collisions STAR does not select events using neutron multiplicity classes \cite{STAR:2023zhao}. Consequently, our calculations use $P(B_\perp) = 1$ for the ultra-peripheral isobar collisions. For the comparison to ALICE data, we will present different neutron multiplicity classes.

\section{Vector meson production at high energy}
\label{sec:vm_production}
As discussed in the introduction, we employ the dipole picture to describe vector meson production in photon-nucleus scattering at high energy. In the nucleus rest frame, the lifetime of a fluctuation of the incoming photon into a quark-antiquark dipole is much longer than the characteristic timescale of the dipole-target interaction. Consequently, the scattering amplitude can be factorized into a convolution of photon and vector meson wave functions and the dipole-target interaction. 
The latter is described in the CGC framework~\cite{Kovchegov:2012mbw,Iancu:2003xm,Gelis:2010nm,Albacete:2014fwa,Morreale:2021pnn}. 
It can be written as \cite{Kowalski:2006hc,Hatta:2017cte,Mantysaari:2020lhf} (see also Refs.~\cite{Mantysaari:2022kdm,Mantysaari:2021ryb,Mantysaari:2022bsp,Boussarie:2016bkq} for recent developments towards NLO accuracy)
\begin{align}
    \widetilde{\Acal}_A(x,\bt) &= 2i\int \dd[2]{\rt}   \frac{\dd{z}}{4\pi}  [\Psi_V^* \Psi_\gamma](\rt,z)  N_A(x,\rt,\bt) \,.
    \label{eq:jpsi_am}
\end{align}
Here, $z$ is the fraction of the large photon lightcone momentum carried by the quark, $\rt$ is the relative transverse spatial separation of the $q\bar q$ dipole, and $\bt$ is the impact parameter measured relative to the target center. The quasi-real photon $\gamma \to q\bar q$ splitting is described by the photon light-cone wave function $\Psi_\gamma$~\cite{Kovchegov:2012mbw}. 
The non-perturbative vector meson wave function $\Psi_V$ is parametrized using the Boosted Gaussian model from~\cite{Kowalski:2006hc}, with model parameters constrained by the decay width data. 

The dipole-target scattering amplitude 
$N_A(x,\rt,\bt)$ describes the eikonal propagation of the quark-antiquark pair in the target color field and reads
\begin{multline}
N_{A}(x,\rt,\bt) =  \\ 1 - \frac{1}{\nc} \tr \left[ V\left(\bt + (1-z) \rt\right) V^\dagger\left(\bt - z \rt \right) \right]\,,
\end{multline}
where $V(\xt)$ is the light-like Wilson line,  which describes the color rotation of a quark state when propagating through the target field at transverse coordinate $\xt$, and it is given by
\begin{equation}
  V(\xt) = \mathrm{P}_{-}\left\{ \exp\left({-ig\int_{-\infty}^\infty \dd{x^{-}} \frac{\rho^a(x^-,\xt) t^a}{\boldsymbol{\nabla}^2 - m^2} }\right) \right\}\,.
  \label{eq:wline_regulated}
\end{equation}
Here, $\mathrm{P}_{-}$ represents path ordering in the $x^-$ direction, and we introduced the infrared regulator $m$, which is needed to avoid the emergence of unphysical Coulomb tails. 
As in \cite{Schenke:2012wb}, the color charges $\rho^a$ (with $a$ the color index) are sampled from a distribution whose width is given by the average squared color charge density, which is obtained from its relation to the local saturation scale extracted from the IPSat dipole-proton amplitude~\cite{Schenke:2012hg}. The geometry of the nucleus and nucleons is essential for determining this saturation scale and will be discussed in detail below.
To simplify the calculation, the Wilson lines are evaluated at a fixed  $x_{1,2}=\sqrt{M_V^2/s}$. When calculating \jpsi production at RHIC kinematics this value would be greater than 0.01, and we compute the Wilson lines at $\xpom=0.01$ to stay in the small-$\xpom$ region where the setup is applicable. 

In this work, we introduce an event-by-event fluctuating nucleon density by following Refs.~\cite{Mantysaari:2016jaz,Mantysaari:2016ykx}. We write the density profile of nucleons $T_p(\bt)$ as
\begin{equation}
\label{eq:Tpfluct}
    T_p(\bt) = \frac{1}{N_q} \sum_{i=1 }^{N_q} p_i T_q(\bt-\bti),
\end{equation}
with the single hot spot density distribution $T_q(\bt) = \frac{1}{2\pi B_q} e^{-{\mathbf b}_\perp^2/(2B_q)}\,$. The coefficients $p_i$ allow for different normalizations for individual hot spots. They follow a log-normal distribution with the width $\sigma$ controlling the magnitude of the density fluctuations. The sampled $p_i$ are normalized to make their expectation value one. 

This prescription results in $N_q$ hot spots with width $B_q$, whose positions $\bti$ are sampled from a two-dimensional Gaussian distribution of width $B_{qc}$. The center of mass is shifted to the origin (of the nucleon) in the end. In this work, we use the Maximum a Posteriori (MAP) parameter set from a Bayesian analysis, in which the geometric parameters of the proton were constrained by the exclusive \jpsi production data from HERA~\cite{Mantysaari:2022ffw} at $x = 1.7 \cdot 10^{-3}$ with $N_q \equiv 3$.

To model the geometric shape of large nuclei, we first sample nucleon positions from a Woods-Saxon distribution 
\begin{equation}\label{eq:WS}
    \rho(r,\theta) = \frac{\rho}{1+\exp[(r-R'(\theta))/a_{\rm WS}]}\,,
\end{equation}
with $R'(\theta)=R_{\rm WS}[1+\beta_2 Y_2^0(\theta)+\beta_3 Y_3^0(\theta) +\beta_4 Y_4^0(\theta)]$, and $\rho$ is the nuclear density at the center of the nucleus. Here $R_{\rm WS}$ is the radius parameter, $a_{\rm WS}$ is the skin diffuseness, and $\theta$ is the polar angle. The spherical harmonic functions $Y_l^m(\theta)$ and the parameters $\beta_i$ account for possible deformations.
Following~\cite{Moreland:2014oya,Schenke:2020mbo}, we further impose a minimal distance of $d_{\rm min}=0.9\,{\rm fm}$ between nucleons when sampling in three dimensions. When a nucleon is added and violates the minimum distance criterion with one or more already sampled nucleons, we re-sample its azimuthal angle $\phi$ to keep the distributions of radial distances and polar angles unchanged~\cite{Moreland:2014oya}.   A random rotation of the entire nucleus is applied after the sampling process. 

\begin{figure}
    \centering
    \includegraphics[width=0.44\textwidth]{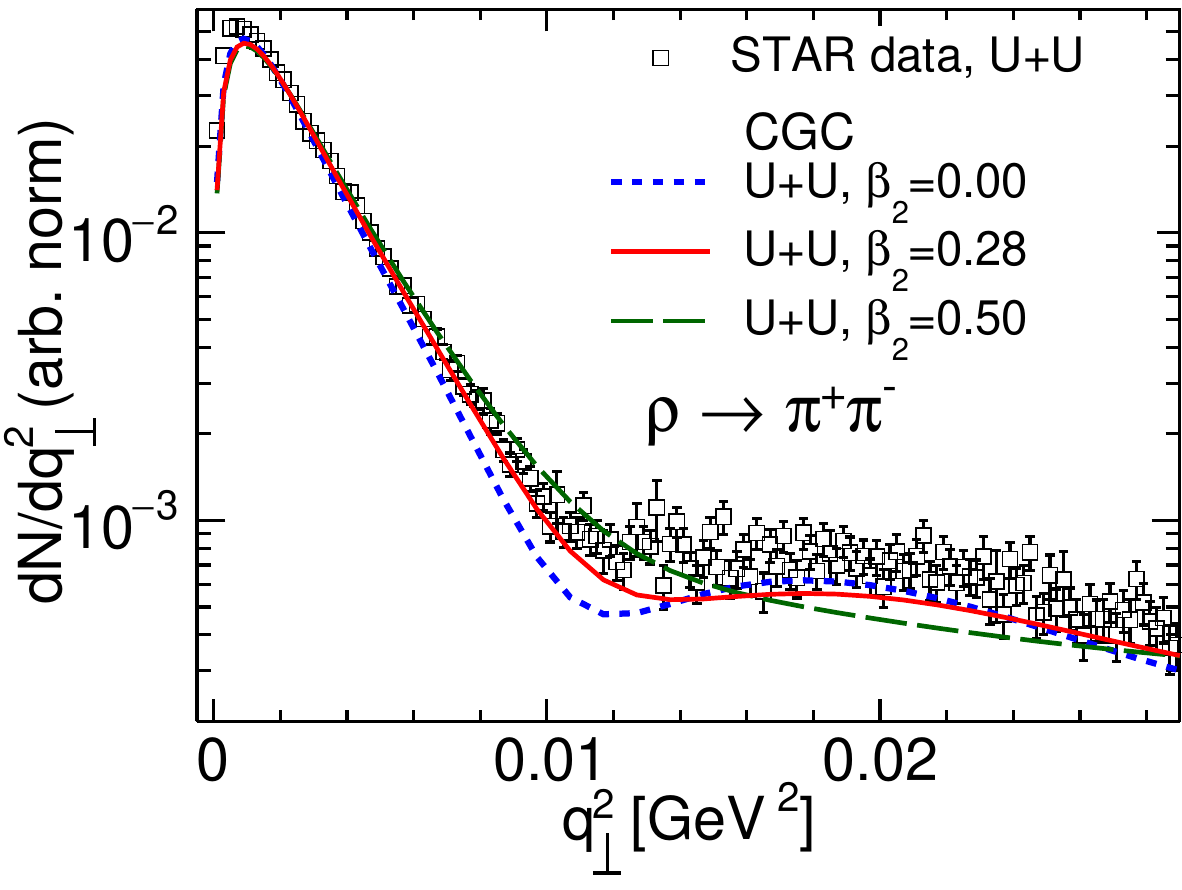}
    \caption{ The $\dd N/\dd q^{2}_{\perp}$ as a function of $q^2_{\perp}$ for  U+U collisions with different $\beta_2$ values averaged over $\Delta \Phi$. The STAR data is from~\cite{STAR:2022wfe}.   }
\label{fig:spectraUU}
\end{figure}

The default Woods-Saxon parameters for gold,  lead, uranium, ruthenium, and zirconium are listed in Table~\ref{tab:isobartable}.   
Note that the radius of uranium is fixed by fitting the STAR data shown in Fig. \ref{fig:spectraUU} under the assumption that $\beta_2 = 0.28$. We note that this value of $R_{\rm WS}=7.2$ fm is close to the mass radius determined from an effective density functional method in \cite{Ryssens:2023fkv}.
We have neglected any deformation for the gold nucleus in this study to provide a cleaner reference for the studies of deformed nuclei. Including a small $\beta_2$ for Au only minimally affects the observables discussed in this paper. In the case of ruthenium, we study the default (case 1) along with three other cases. These cases gradually approach the parameter values for zirconium \cite{Bhatta:2023cqf,Jia:2021oyt,Zhang:2021kxj}.

\begin{table}[t]
  \centering
  \caption{The parameter sets for the nuclear geometries. The parametrizations for Ru and Zr are from Refs.~\cite{Bhatta:2023cqf,Jia:2021oyt,Zhang:2021kxj}}
  \begin{tabular}{llllll}\hline \hline
    Nucleus & $R_{\rm WS}$ [fm] & $a_{\rm WS}$ [fm] & $\beta_2$ & $\beta_3$ & $\beta_4$\\ \hline
    Au & 6.38 & 0.54 &  0.0 & 0.0 & 0.0 \\ \hline
    Pb & 6.62 & 0.546 &  0.0 & 0.0 & 0.0 \\ 
    \hline
    U & 7.20 & 0.55 &  0.28 & 0.0 & 0.093 \\ 
    \hline
    Ru (case 1) & 5.09 & 0.46 &  $0.16$ & 0.0 & 0.0 \\ 
    Ru (case 2) & 5.09 & 0.46 &  $0.16$ & 0.20 & 0.0 \\ 
    Ru (case 3) & 5.09 & 0.46 &  $0.06$ & 0.20 & 0.0 \\ 
    Ru (case 4) & 5.09 & 0.52 &  $0.06$ & 0.20 & 0.0 \\ \hline
    Zr & 5.02 & 0.52 &  0.06 & 0.20 & 0.0 \\ \hline
     \hline
  \end{tabular}
  \label{tab:isobartable}
\end{table}

\begin{figure}
    \centering
    \includegraphics[width=0.44\textwidth]{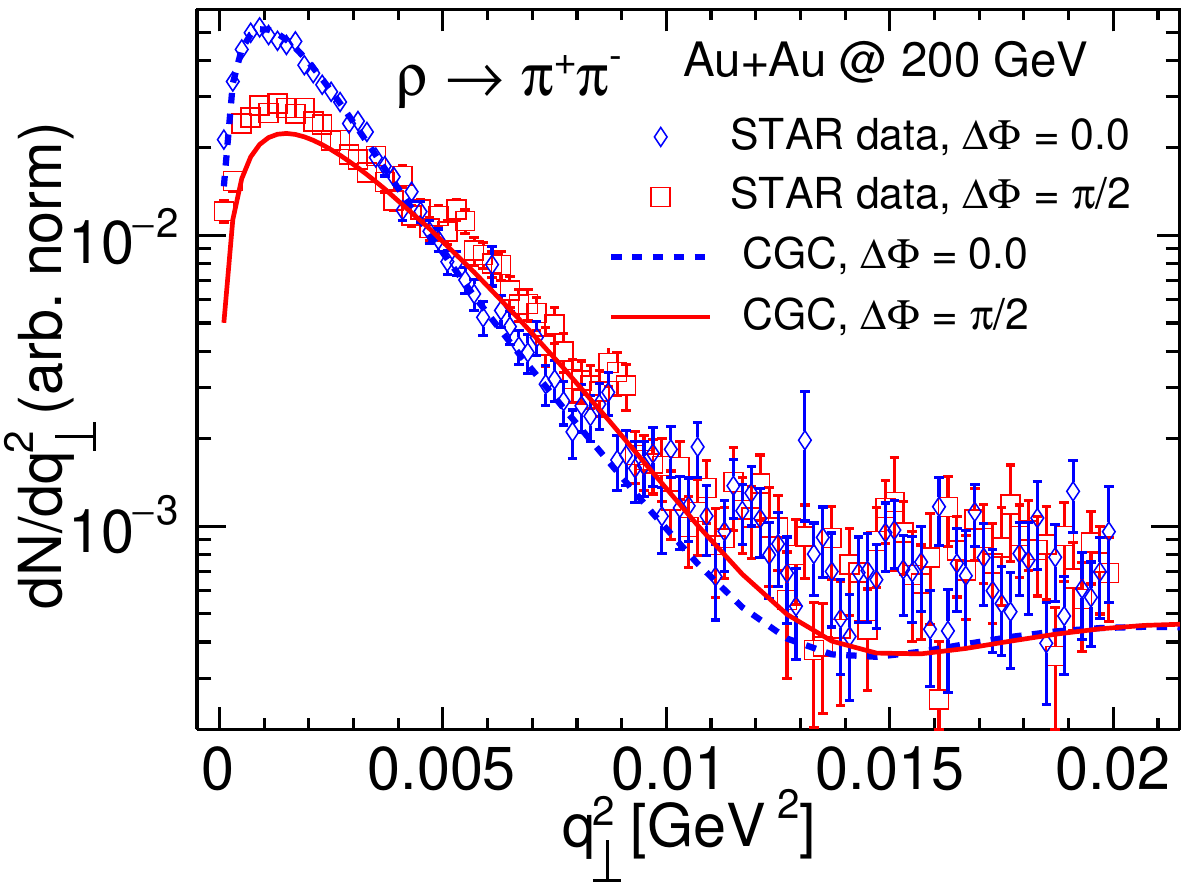}
    \caption{ The $\dd N/\dd q^{2}_{\perp}$ as a function of $q^2_{\perp}$ for Au+Au collisions  for $\phi$ bins at $0^\circ$ and $90^{\circ}$. The STAR data is from~\cite{STAR:2022wfe}.   }
\label{fig:spectraAuAu}
\end{figure}

\begin{figure}
    \centering
    \includegraphics[width=0.44\textwidth]{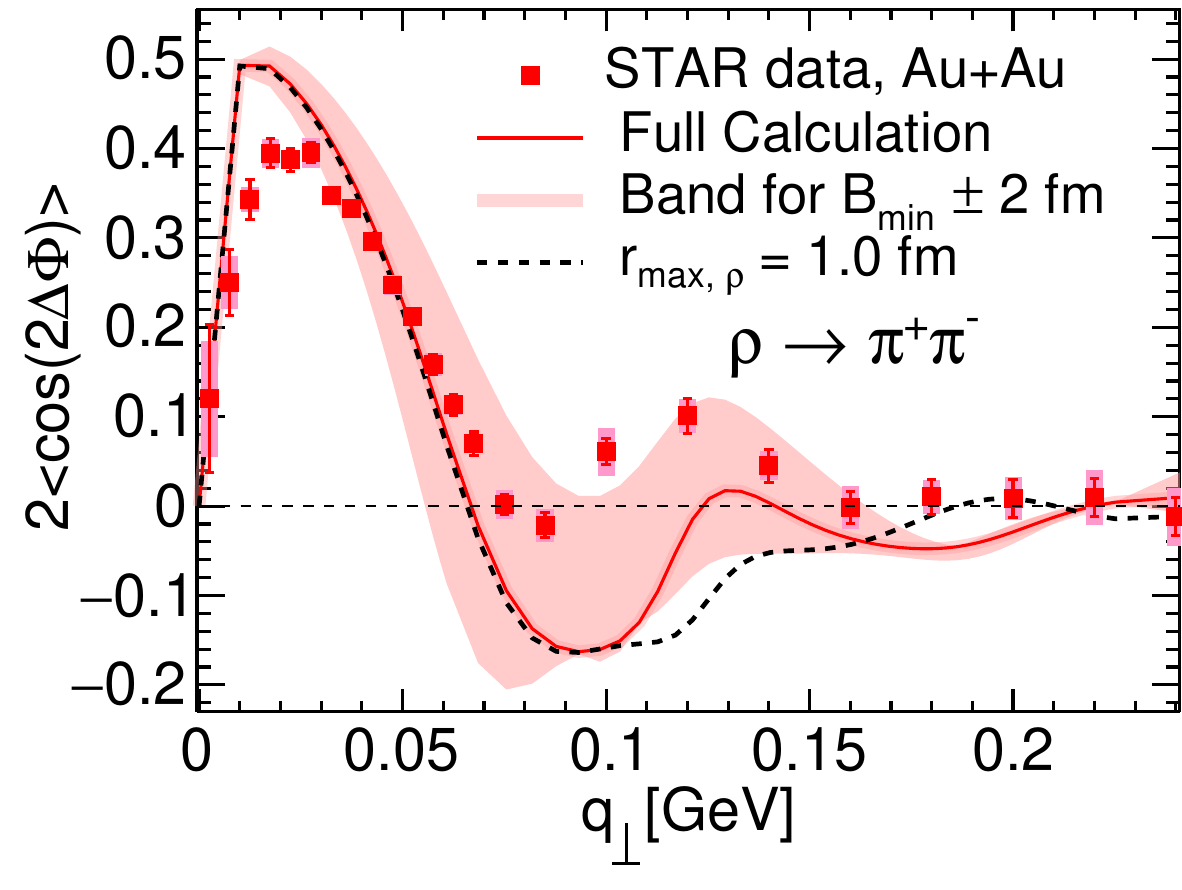}
    \caption{ The  $2\left<\cos2\Delta\Phi\right>$ modulation as a function of $q_{\perp}$ of $\rho\rightarrow\pi^{+}\pi^{-}$ in Au+Au collisions. The red band is calculated by varying the $B_{\rm min}\pm 2$ fm: the upper limit corresponds to smaller $B_{\rm min}$ and the lower limit to larger $B_{\rm min}$.  The black line is calculated with the wave function of $\rho$ cut at $r_{\rm max, \rho}$ = 1.0 fm. The STAR data is from~\cite{STAR:2022wfe}. 
    }
\label{fig:v2pt}
\end{figure}

\begin{figure}
    \centering
    \includegraphics[width=0.44\textwidth]{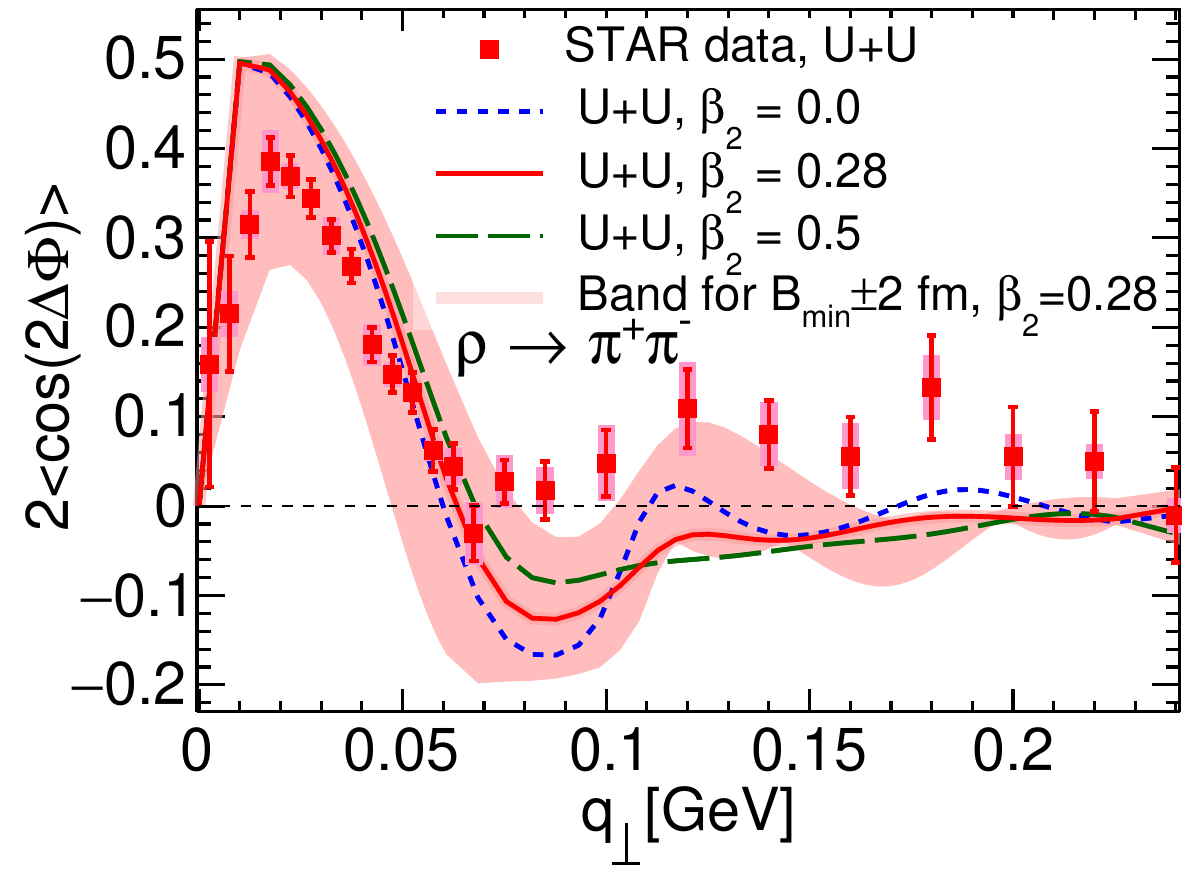}
    \caption{The $2\langle \cos 2\Delta\Phi\rangle$ modulation as a function of $q_\perp$ for $\rho \to \pi^+\pi^-$ in U+U collisions with different deformation parameter $\beta_2$. The red band is calculated by varying the $B_{\rm min}\pm 2$ fm: the upper limit corresponds to smaller $B_{\rm min}$ and the lower limit to larger $B_{\rm min}$ for $\beta_2 = 0.28$.  The STAR data is from~\cite{STAR:2022wfe}.}
    \label{fig:beta2_qdep}
\end{figure}

\begin{figure}
    \centering
    \includegraphics[width=0.45\textwidth]{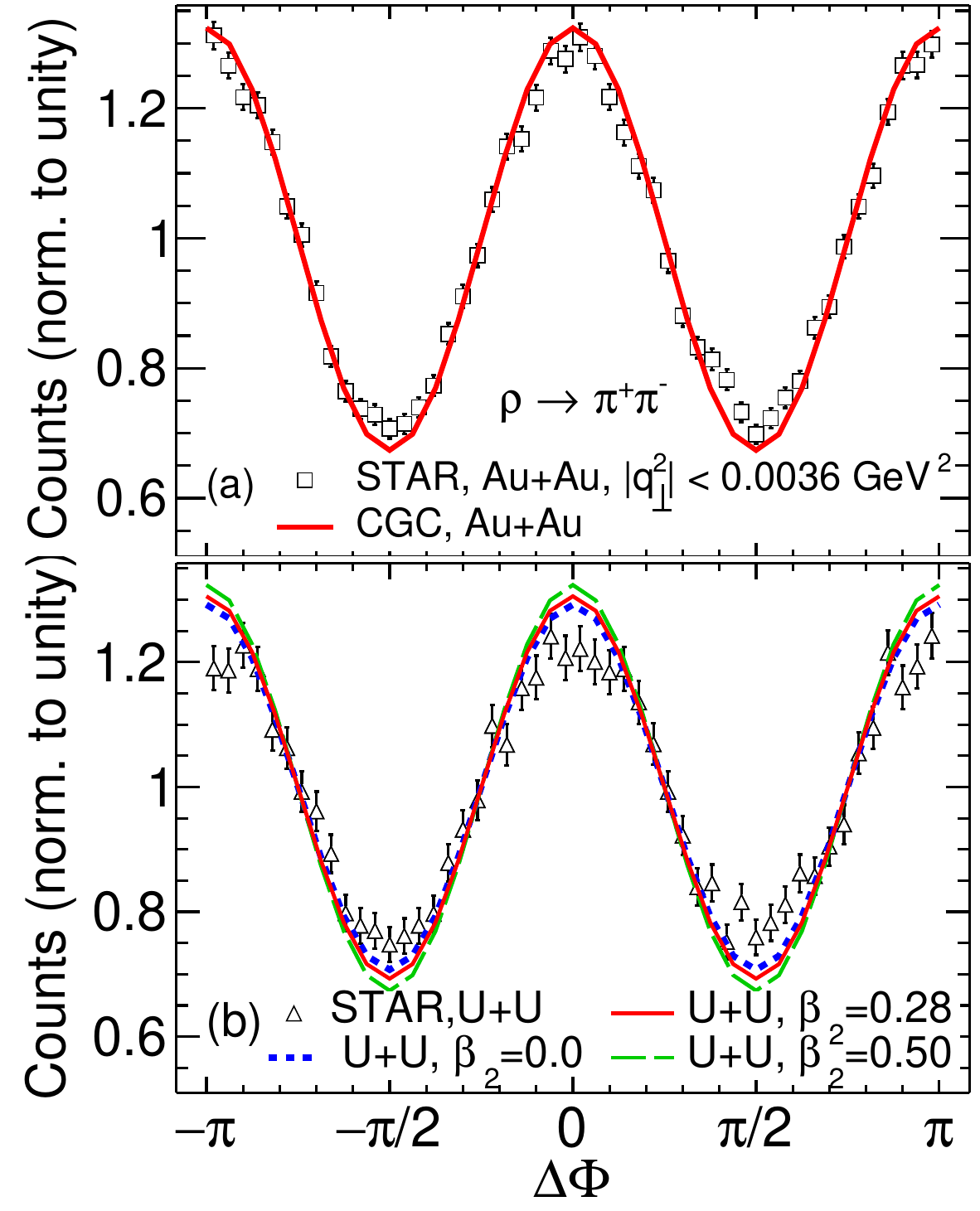}
    \caption{ The $\Phi_P - \Phi_q$ distribution for $\pi^+\pi^-$ pairs collected from Au+Au (a) and U+U collisions with different $\beta_2$ values (b) with a pair transverse momentum less than 60 MeV. The STAR data is  from~\cite{STAR:2022wfe}.}
\label{fig:integratedv2}
\end{figure}

\section{RESULTS}\label{sec:results}
\subsection{ $\rho$ production in ultra-peripheral Au+Au and U+U collisions}
\label{sec:vmprodinAuAuUU}

The transverse momentum squared, $q_\perp^2$, dependence of $\rho$ meson production in ultra-peripheral Au+Au collisions for the pair rapidity $\vert y \vert < 1.0$ and pair 
invariant mass of $0.65 < Q < 0.9$ GeV, as measured by the STAR Collaboration~\cite{STAR:2022wfe} and calculated in our framework, is shown in Fig.~\ref{fig:spectraAuAu}. As the STAR data is presented with arbitrary normalization, we normalize the calculated spectra such that we reproduce the measured cross-section integrated over $q_\perp < 60$ MeV at $\Delta\Phi=0$. 

The differential cross-section vanishes as $q_\perp \to 0$ as a result of the destructive interference. 
Turning our attention to the angular dependence, after reaching a peak, the spectrum at $\Delta\Phi = 0$ decreases faster as a function of $q_\perp^2$ than the spectrum at $\Delta\Phi = \pi/2$. 
For our calculation to reproduce all these features, we must consider the linear polarization of photons, the interference effects, and the photon $\kt$ \cite{Mantysaari:2022sux}.
We note that the slope of the $q_\perp^2$ spectrum is sensitive to both the nuclear and vector meson size and as such depends on the $\rho$ wave function in the poorly constrained region of non-perturbatively large dipoles. However, the $\rho$ meson size is mainly determined by its mass, and all realistic $\rho$ wave function models are expected to result in very similar sizes. 

In Figure~\ref{fig:v2pt}, we present the  $2\left\langle\cos2\Delta\Phi\right\rangle$ modulation as a function of $q_{\perp}$ in ultra-peripheral Au+Au collisions, with the solid red line representing the full calculation. The band indicates the systematic uncertainty resulting from the lack of precise knowledge of the lower limit of the inter-nuclear impact parameter, $B_{\rm min}$.
The upper and lower bounds of the band shown in Fig.\,\ref{fig:v2pt} correspond to the results when $B_{\rm min}$ is decreased or increased by 2 fm, respectively. 

Our default setup, where $B_{\rm min}$ is determined by ensuring $N_\mathrm{coll}=0$ as obtained from the Monte-Carlo Glauber model, results in $\langle B_{\rm min} \rangle=15$ fm. This means that the scenario with reduced $B_{\rm min}$ leads to an average value $\left\langle B_{\rm min} \right\rangle\approx 13$ fm, which is close to $2R_{\rm WS}$, the value used in \cite{Xing:2020hwh}. Reassuringly, the result corresponding to this choice for $B_{\rm min}$, the upper limit of the shaded band, is consistent with that of Ref.~\cite{Xing:2020hwh}, which uses exactly $B_{\rm min}=2R_{\rm WS}$. This exercise demonstrates the important effect that the minimal distance between the nuclei has on the angular modulation. It is expected that interference effects vanish at very large impact parameters, as they disappear at very large $q_\perp$ \cite{Mantysaari:2022sux,Bertulani:2005ru}. Reversely, interference effects, in this case, angular modulations, are stronger for small impact parameters and are thus increased when decreasing $B_{\rm min}$. Fig.\,\ref{fig:v2pt} also shows that the variation of $2\left\langle\cos2\Delta\Phi\right\rangle$ with $B_{\rm min}$ is mild at low $q_\perp$ and becomes maximal around $q_\perp\approx 0.1\,{\rm GeV}$. This is because variations in $B_{\rm min}$ need to be at least of order $1/q_\perp$ to show a significant effect.

To study the sensitivity of our results to non-perturbatively large distance scales we also show in Fig.~\ref{fig:v2pt} the $2\left\langle\cos2\Delta\Phi\right\rangle$ modulation calculated by neglecting the contribution from dipole sizes $|\rt| >1$ fm in Eq.~\eqref{eq:jpsi_am}. In this case, the obtained modulation is identical up to $q_\perp \approx 0.1$ GeV. In the larger transverse momentum region there is moderate sensitivity to non-perturbatively large dipoles, and we consider our results to be most reliable in the $q_\perp<0.1$ GeV region.

Fig.~\ref{fig:spectraUU}  shows the effects of nuclear deformation on the $q^2_{\perp}$ dependence of the (angle averaged) spectra in ultra-peripheral U+U collisions.
As already mentioned in Sec.~\ref{sec:vm_production}, we determine an optimal value for the uranium radius parameter $R_\mathrm{WS}$ by requiring that the $q_\perp$ dependence of the STAR data is reproduced. 
It is evident that increasing the degree of deformation leads to a flatter $\dd N/\dd q^2_{\perp}$ distribution. The sensitivity to deformations is due to two separate effects. First, a larger $\beta_2$ value results in increased fluctuations that enhance the incoherent cross-sections, causing the total cross-section to drop at a slower rate at large $q_\perp^2$ compared to cases with smaller $\beta_2$~\cite{Mantysaari:2023qsq}. Second, the larger $\beta_2$ value leads to a wider $B_{\rm min}$ distribution. Configurations that allow for small $B_{\rm min}$, that have a different degree of interference and larger photon transverse momentum, result in flatter spectra and enter with a larger weight~\cite{Mantysaari:2022sux}. 

To demonstrate the sensitivity of the angular modulation to the deformed structure of uranium we show in Fig.~\ref{fig:beta2_qdep} the extracted magnitude of the modulation, $2\langle \cos 2\Delta \Phi \rangle$, as a function of $\rho$ meson transverse momentum. Strong sensitivity on $\beta_2$ is seen especially in the $0.05\,\mathrm{GeV}<q_\perp < 0.1\,\mathrm{GeV}$ region.\footnote{We emphasize that, as demonstrated above, results in this kinematical domain are not sensitive to non-perturbatively large dipoles.}
Again, the main driver of the $\beta_2$ dependence is the change in impact parameter distributions with $\beta_2$. As mentioned in our discussion of Fig.\,\ref{fig:v2pt}, to achieve a noticeable effect on the modulation, changes in $B_{\rm min}$ need to be of order $1/q_\perp$, explaining why we do not observe strong effects of different $\beta_2$ at small $q_\perp$.
The positive modulation coefficient measured for $q_\perp \gtrsim 0.1\,\mathrm{GeV}$ is only reproduced for reduced $B_{\rm min}$, shown as the upper end of the band for the $\beta_2=0.28$ case. 

Fig.~\ref{fig:integratedv2} shows the $\Delta\Phi$ distribution for Au+Au and U+U events. Here, we include the $\pi^-\pi^+$ pairs with a pair transverse momentum squared $q^2_{\perp}< 0.0036~{\rm GeV^2}$, corresponding to the range used by the STAR Collaboration~\cite{STAR:2022wfe}. 
Each distribution is scaled such that the average yield (integrated over $\Delta\Phi$) is normalized to unity.
Both the Au+Au and U+U data sets exhibit a clear and prominent $\cos(2\Delta\Phi)$ modulation, which is well reproduced by the CGC calculations. 
The $\beta_2$ dependence seen in Fig.\,\ref{fig:beta2_qdep} is present here, but it is small in the chosen $q_\perp^2$ range. As can already be seen from the $q_\perp^2$ spectra shown in Fig.~\ref{fig:spectraUU}, the cross-section is sensitive to deformation only in the larger $q_\perp^2 \gtrsim 0.006\,\mathrm{GeV}^2$ region.

\begin{figure}
    \centering
    \includegraphics[width=0.44\textwidth]{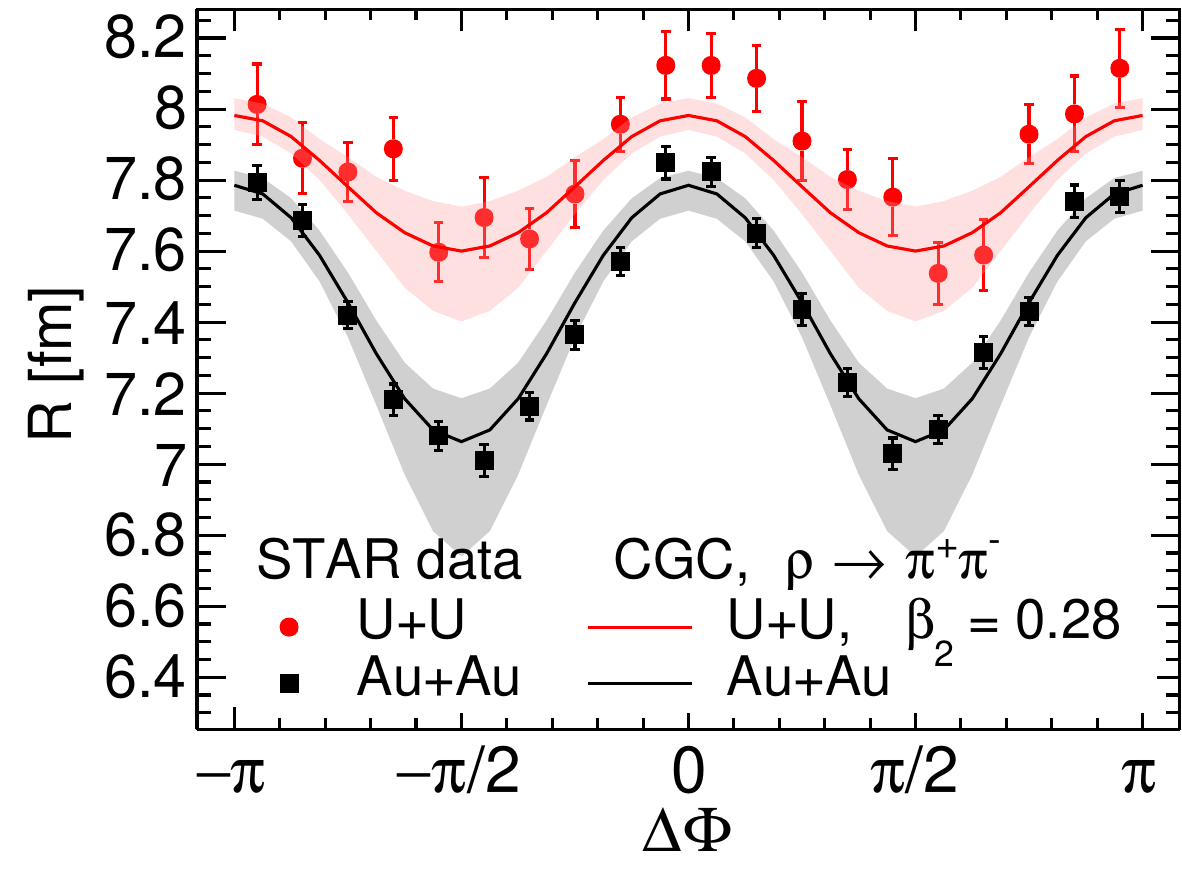}
    \caption{The effective radius obtained from a fit using Eq.\,\eqref{eq:fit_ws} as a function of the  $\Phi_P - \Phi_q$ angle for Au+Au and U+U with $\beta_2=0.28$.
    The bands show errors from varying $B_{\rm min}$ by 2 fm along with errors from the fit used to extract $R$. The lower edge of the band corresponds to the smaller $B_{\rm min}$.
    The STAR data is from~\cite{STAR:2022wfe}.
    }
\label{fig:RAuAuUU}
\end{figure}

\begin{figure}
    \centering
    \includegraphics[width=0.44\textwidth]{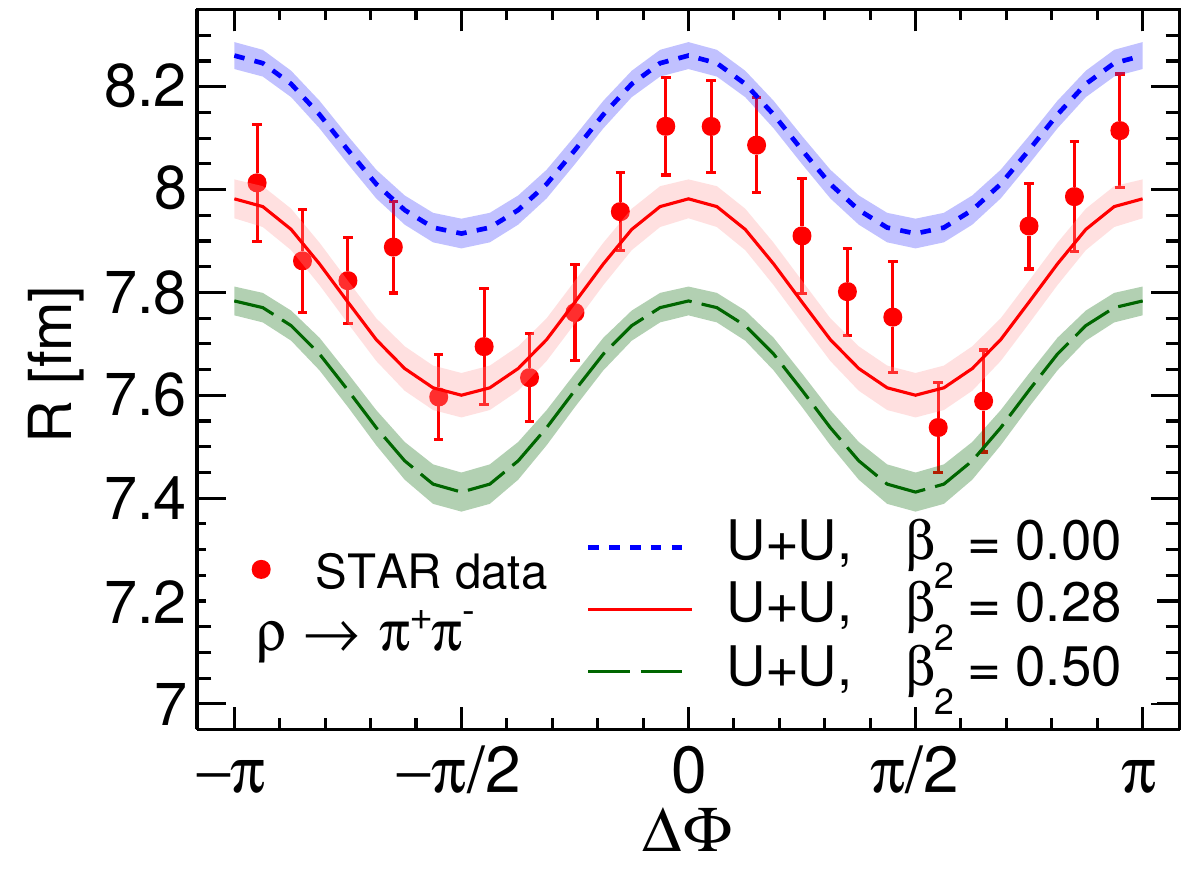}
    \caption{The effective radius obtained from a fit using Eq.\,\eqref{eq:fit_ws} as a function of the  $\Phi_P - \Phi_q$ angle for U+U with  different $\beta_2$ values.
    The bands indicate the uncertainties from extracting the radius.
    The STAR data is from~\cite{STAR:2022wfe}.}
\label{fig:RUU}
\end{figure}

The STAR collaboration has also extracted the nuclear radius by using an empirical model to fit the measured $q^2_{\perp}$-spectrum~\cite{STAR:2017enh,STAR:2022wfe}.  This model is composed of a coherent contribution characterized by the form factor of a spherically symmetric Woods-Saxon distribution, Eq.~\eqref{eq:WS},  and an incoherent contribution characterized by a dipole form factor. 
Following the STAR method, we perform the Fourier transform of the thickness function obtained from the density distribution in Eq.~\eqref{eq:WS} to get the nuclear form factor, 
denoted as $F[...](q^2_{\perp})$ in Eq.~\eqref{eq:fit_ws} below (see also Ref.~\cite{Mantysaari:2022sux} for a discussion on how the saturation effects modify the nuclear geometry). The effective radius is then extracted by fitting the calculated $q^2_{\perp}$-spectra with a function of the form
\begin{equation}
    \label{eq:fit_ws}
    f(q^2_{\perp}) = A_{c}\lvert F\left[ \rho_A(r; R, a) \right](|q^2_{\perp}|) \rvert^{2} + \frac{A_{i}/Q_0^2}{ (1 + |q^2_{\perp}|/Q_0^2)^2 }.
\end{equation}
Here $A_c$ is the amount of coherent production and $A_{i}$ is the amount of incoherent production. 
We use the value $Q_0^2=0.099$ ${\rm GeV}^{2}$ for $\rho$ , $Q_0^2=0.084$ ${\rm GeV}^{2}$ for $\phi$  and $Q_0^2=0.53$ ${\rm GeV}^{2}$ for \jpsi production obtained by fitting the calculated $e$+A incoherent photoproduction cross-sections at $0.2<\vert t\vert<0.5~{\rm GeV^2}$. 
Then, we determine  optimal values for the  parameters $A_c$, $A_i$ and $R$ in  Eq.~\eqref{eq:fit_ws} by fitting the calculated $q^2_\perp$ spectra within $0.0034<q^2_\perp < 0.0127~{\rm GeV^2}$. 

\begin{figure*}
    \centering
    \includegraphics[width=0.95\textwidth]{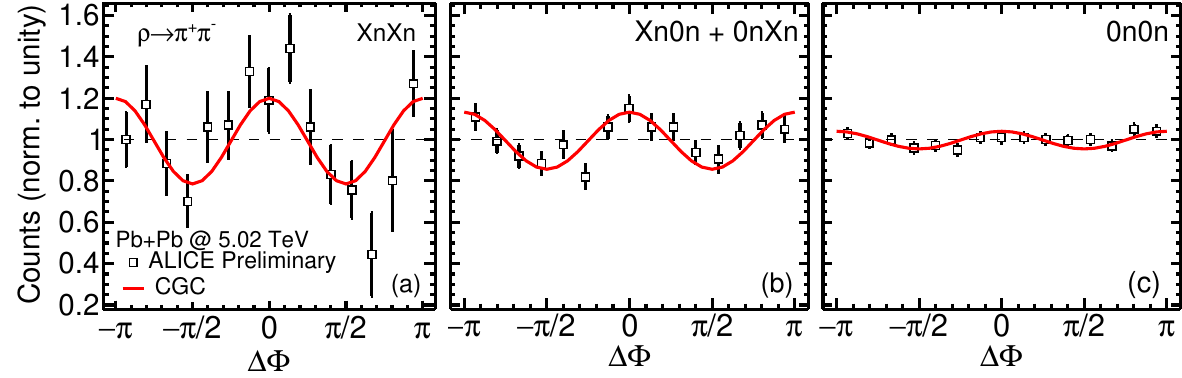}
    \caption{ The $\Delta\Phi = \Phi_P - \Phi_q$ distribution for $\pi^+\pi^-$ pairs collected from Pb+Pb at 5.02 TeV with XnXn (a) Xn0n + 0nXn (b) and 0n0n (c) with a pair transverse momentum less than 100 MeV. The preliminary ALICE data is from~\cite{ALCIE:2023riffero}.}
\label{fig:rholhc}
\end{figure*}

Fig.~\ref{fig:RAuAuUU} presents the fitted radius $R$ as a function of $\Delta\Phi$ in Au+Au and U+U collisions computed using the parameters provided in Table~\ref{tab:isobartable}. The plot reveals a clear modulation in the resulting $R$ values with respect to $\Delta\Phi$, similar to the behavior observed in Fig.~\ref{fig:integratedv2}. 
Consistent with the STAR data, U+U collisions show a weaker $\cos(2\Delta\Phi)$ modulation than Au+Au collisions. This difference can be attributed to the larger radius of uranium nuclei, which leads to larger $B_{\rm min}$ values. As discussed above, larger $B_{\rm min}$ reduces interference effects and leads to smaller modulation. This effect is also seen in the systematic error band shown in this figure, where again we have varied $B_{\rm min}$ by $\pm 2$ fm. The lower edge of the band, which varies more strongly with $\Delta\Phi$ than the upper edge, corresponds to the smaller overall $B_{\rm min}$.

It is noteworthy that the fitted radii $R$ are significantly larger than the input parameters $R_{\rm WS}$ for the Woods-Saxon distribution 
shown in Table~\ref{tab:isobartable}. This difference is due to the contributions from the finite size of the $\rho$ meson wave function, interference effects, and the photon transverse momentum (with the latter leading to smaller effective radii), and as such $R$ should not be directly interpreted as the nuclear radius. 

To remedy this situation, we also extract the radii of gold and uranium from our calculations using the same method as STAR~\cite{STAR:2022wfe}, which estimates the contributions from the interference effect and the finite size $\rho$ wave function. We find ${\rm R_{Au} = 6.59}\pm 0.02$ fm, and ${\rm R_{U} = 7.31\pm0.04}$ fm, which are consistent with the STAR data~\cite{STAR:2022wfe}.
The extracted radius from our calculations of $^{197}{\rm Au}$, which use the commonly used default value for $R_{\rm WS}$ is consistent with the 
world measurements of Au charge radii plus neutron skin width at low energies, with ${\rm R_{Au}} = 6.45\pm0.27$ fm from the German Electron Synchrotron (DESY) and ${\rm R_{Au}} = 6.74\pm0.06$ fm from Cornell \cite{Mcclellan:1971rxw,DeVries:1987atn,Centelles:2008vu,PREX:2021umo}.
As we have chosen an $R_{\rm WS}$ for uranium to be larger than the commonly used value to fit the spectra, our result for the radius of $^{238}{\rm U}$ agrees with the STAR result and
is larger than that expected from its charge radius ${\rm R_{U}} = 6.90\pm0.06$ fm measured at DESY plus the known value of neutron skins of similar nuclei \cite{Mcclellan:1971rxw,DeVries:1987atn,Centelles:2008vu,PREX:2021umo}.

To further investigate the effect of nuclear deformations on the effective radius $R$, we present the radius as a function of $\Delta \Phi$  using different deformation parameters $\beta_2=0.0, 0.28$ and 0.50 in Fig.~\ref{fig:RUU}.
Smaller deformations (small $\beta_2$) are found to result in larger effective radii $R$, a trend consistent with the results in Fig.~\ref{fig:spectraUU} where a steeper spectrum corresponding to a larger size is obtained with smaller $\beta_2$.
See discussion of Fig. \ref{fig:spectraUU} for more details. We note that the stronger dependence on $\beta_2$ compared to what is observed in Fig.~\ref{fig:integratedv2} is a result of including here contributions from higher $q_\perp^2$, which are sensitive to shorter scale structures.

To complete this section, we present in Fig.~\ref{fig:rholhc} the angular modulation in $\rho \to \pi^+ \pi^-$ production in Pb+Pb collisions at the LHC compared with the preliminary ALICE data~\cite{ALCIE:2023riffero}. At the LHC energy, the nucleus is probed at much smaller Bjorken-$x$ values compared to RHIC, where the nuclear saturation scale is larger. This further suppresses contributions from the non-perturbatively large dipoles. 
The ALICE data is reported separately in different neutron multiplicity classes \cite{ALCIE:2023riffero}. Fig. \ref{fig:rholhc} shows the $\Delta\Phi$ distribution for Pb+Pb collisions at $\sqrt{s}=5.02$ TeV. Following the ALICE measurement \cite{ALCIE:2023riffero}, we include the $\pi^-\pi^+$ pairs with a pair invariant mass of $0.6 < Q < 0.95$ GeV, a pair transverse momentum $q_{\perp}< 0.1~{\rm GeV}$ and rapidity values within -0.8 and 0.8. 
Each distribution is scaled so that the average yield, integrated over $\Delta\Phi$, is normalized to unity.
The magnitude of the modulation strongly depends on the neutron multiplicity class, with the most pronounced modulation observed in the ``XnXn" case. This dependence arises from the fact that events with higher forward neutron multiplicities are biased towards smaller impact parameters. As discussed earlier, smaller nucleus-nucleus distances result in an enhanced angular modulation. Our calculation accurately reproduces this dependence on neutron multiplicity.

\begin{figure}
    \centering
    \includegraphics[width=0.44\textwidth]{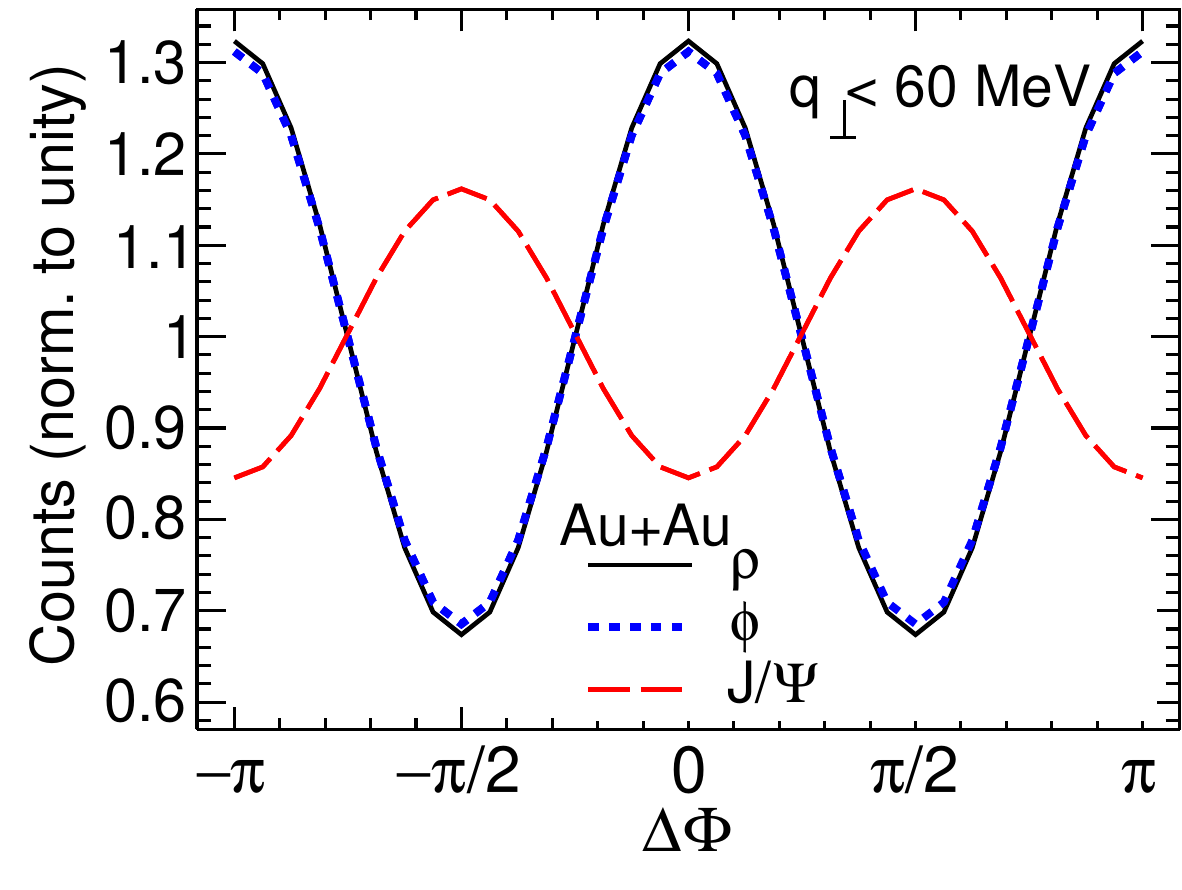}
    \caption{ The $\Phi_P - \Phi_q$ distribution for $\rho\rightarrow\pi^+\pi^-$, $\phi\rightarrow K^+K^-$ and $\jpsim \rightarrow e^+e^-$ pairs collected from Au+Au at 200 GeV with a pair transverse momentum less than 60 MeV. 
    }
\label{fig:v2phijpsi}
\end{figure}

\begin{figure}
    \centering
    \includegraphics[width=0.44\textwidth]{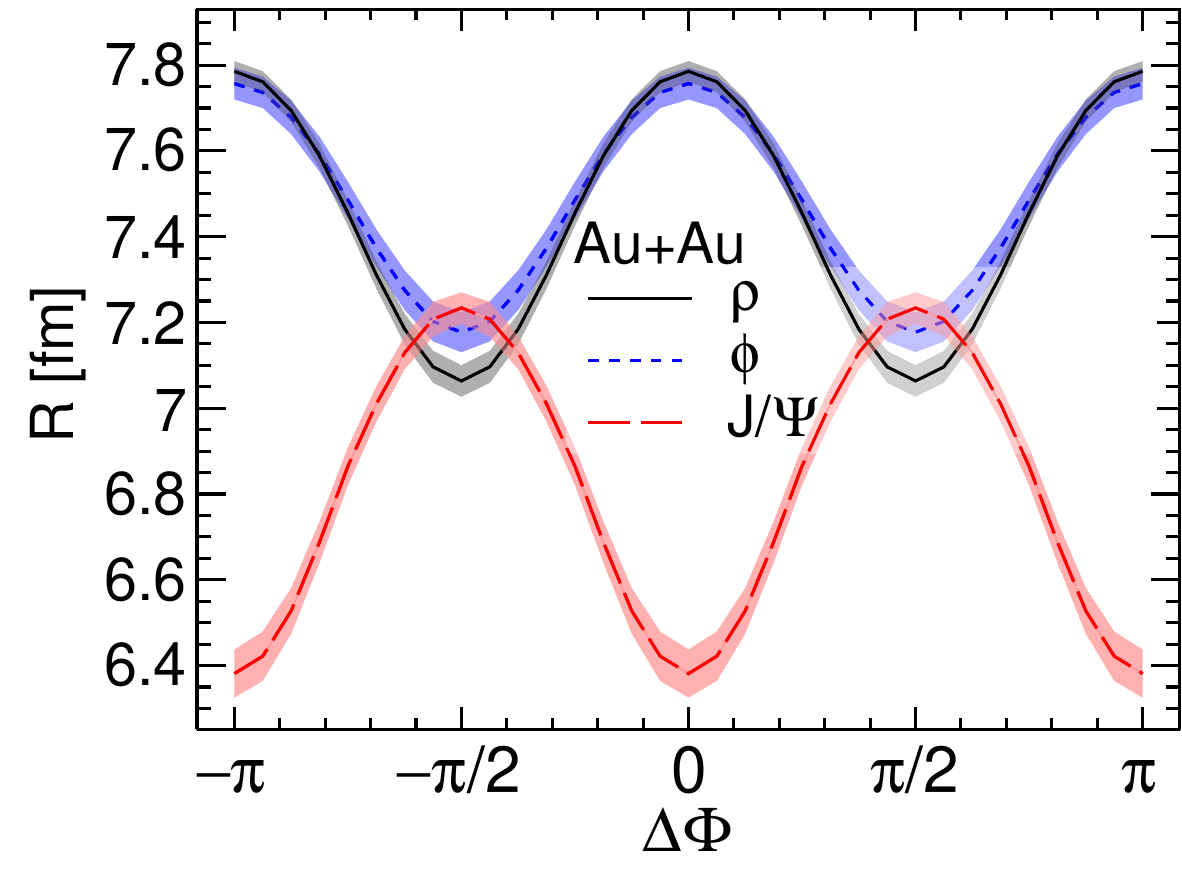}
    \caption{The effective radius obtained from a fit using Eq.\,\eqref{eq:fit_ws} as a function of the  $\Phi_P - \Phi_q$ angle for $\rho\rightarrow\pi^+\pi^-$, $\phi\rightarrow K^+K^-$ and $\jpsim \rightarrow e^+e^-$  at Au+Au 200 GeV.
    The bands indicate the fit uncertainties.}
\label{fig:Rv2phijpsi}
\end{figure}

\begin{figure}
    \centering
    \includegraphics[width=0.44\textwidth]{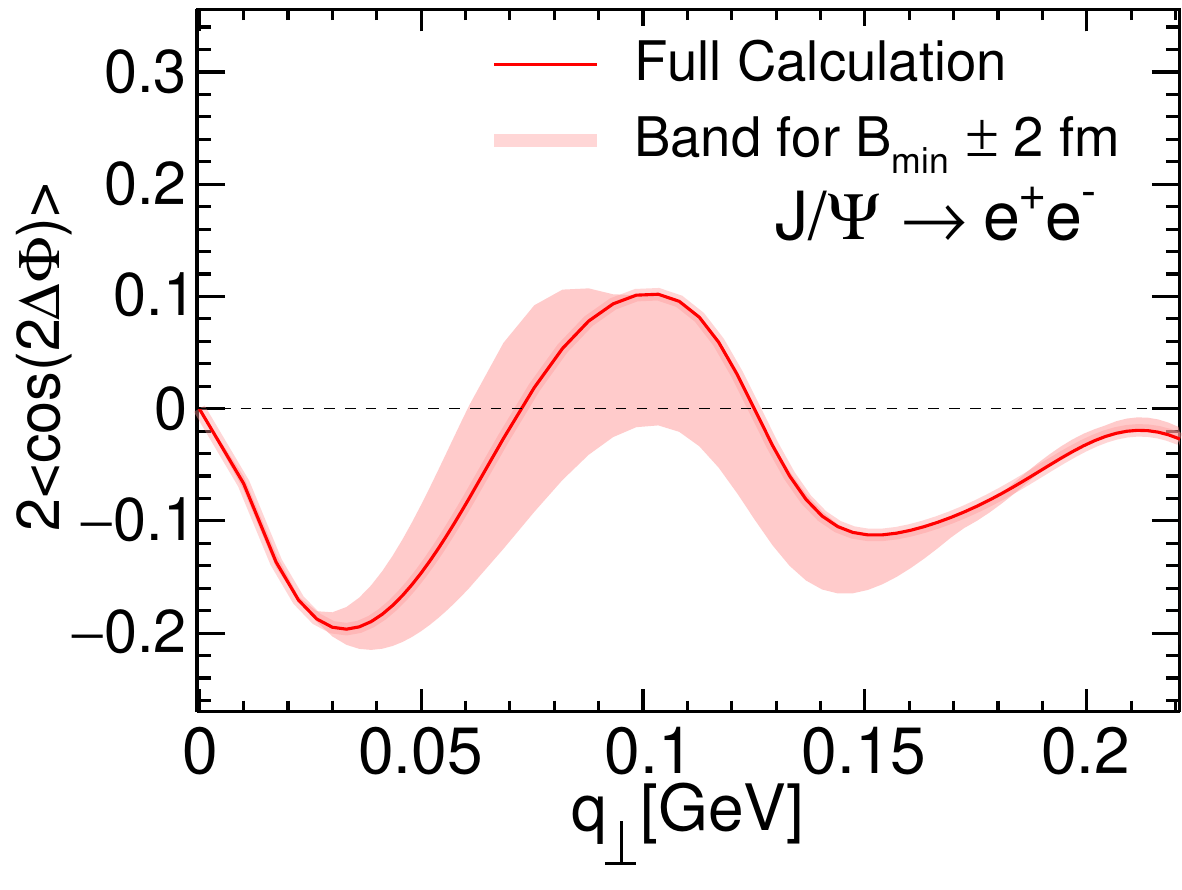}
    \caption{ The  $2\left<\cos2\Delta\Phi\right>$ modulation as a function of $q_{\perp}$ of $\jpsim \rightarrow e^{+}e^{-}$ in Au+Au collisions. The band is calculated by varying the $B_{\rm min}\pm 2$ fm. }
\label{fig:v2ptjpsi}
\end{figure}

\subsection{$\phi$ and \jpsi production}
\label{sec:phijpsi}

We now examine the production of vector mesons $\phi$ and \jpsi in ultra peripheral collisions of Au+Au nuclei at 200 GeV. 
The \jpsi production is especially intriguing, as the large mass provides a scale suppressing contributions from the non-perturbatively large dipoles even if the saturation scale of the nucleus is not clearly in the perturbative domain. 

The dependence on the vector meson species is illustrated in Fig.\,\ref{fig:v2phijpsi}, where we present the  $\Delta\Phi = \Phi_P - \Phi_q$ distributions of $\rho\to\pi^+\pi^-$, $\phi\to K^+K^-$, and $\jpsim \to e^+e^-$ production, with the pair transverse momentum being less than 60 MeV.
The invariant mass range for the decay process $\phi\to K^+K^-$ is specified as $1.0<Q<1.1$ GeV. As the \jpsi decay width $\Gamma_{\jpsim} = 5.5\times 10^{-6}$ GeV is very small, we approximate the Breit-Wigner distribution by a delta function when calculating \jpsi production.

We demonstrate that both the $\rho$ and $\phi$ mesons exhibit the same sign for the ${\cos(2\Delta\Phi)}$ modulation, with their magnitudes being remarkably close, which is expected based on the similar form of $\rho$ and $\phi$ wave functions used. The equal sign of the modulation is a result of the decay products, namely $K$ and $\pi$, being both scalar particles. In contrast, for the $\jpsim \rightarrow e^-+e^+$ decay, where the daughter particles are spin-1/2, the ${\cos(2\Delta\Phi)}$ modulation peaks at $\Delta\Phi = \pm \pi/2$, see also Refs.~\cite{Hagiwara:2020juc,Brandenburg:2022jgr}. This sign difference is already apparent from Eq.\,\eqref{eq:modulation}. We note that the modulation of the effective radius with $\Delta\Phi$ is larger for \jpsi production compared to the other mesons. This is in contrast to the smaller modulation of the cross-section for \jpsi shown in Fig.\,\ref{fig:v2phijpsi}. The difference could emerge because the effective radius is sensitive to larger $q_\perp$ compared to those included in Fig.\,\ref{fig:v2phijpsi}.

In Fig.~\ref{fig:Rv2phijpsi}, we present the extracted radii as a function of the angle $\Delta\Phi$ for $\rho$, $\phi$, and \jpsi production. Again, it is evident that the $\rho$ and $\phi$ mesons exhibit similar $\Delta\Phi$ dependencies, whereas the \jpsi meson exhibits the opposite sign for the modulation. We note that the extracted radius in the case of $\phi$ production is slightly smaller than that for $\rho$ meson production. This difference can be attributed to the narrower width of the wave function of the  $\phi$ meson compared to that of the $\rho$ meson, as discussed in Ref.~\cite{Kowalski:2006hc}. 
Additionally, due to the comparatively narrower wave function of the \jpsi vector meson, its extracted radius is the smallest among the three. 

To conclude the discussion of heavy vector meson production, we present the $q_\perp$ dependence of the $\cos 2\Delta\Phi$ modulation in \jpsi production. The predicted modulation in STAR kinematics is shown in Fig.~\ref{fig:v2ptjpsi}. Again, the sensitivity to the uncertainty in $B_\mathrm{min}$ is illustrated by varying the minimum impact parameter by $\pm 2$ fm. As seen before, for \jpsi production, the modulation has the opposite sign from light meson production studied previously, and the magnitude of this modulation is smaller by approximately a factor of $2$. 
We note that, unlike in Ref.~\cite{Brandenburg:2022jgr}, we have not incorporated the correction for soft photon radiation in the case of \jpsi production. This correction predominantly affects results at $q_{\perp} > 0.12~{\rm GeV/c}$ \cite{Brandenburg:2022jgr} and as such our main results should not be significantly modified by this contribution.

\subsection{Vector meson production in ultra-peripheral isobar collisions}
\label{sec:vmprodisobar}
In this section, we present the results for ultra-peripheral \RuRu\ and \ZrZr\ collisions, showcasing the sensitivity of vector meson production to the nuclear structure of the isobars. 

\begin{figure}
    \centering
    \includegraphics[width=0.44\textwidth]{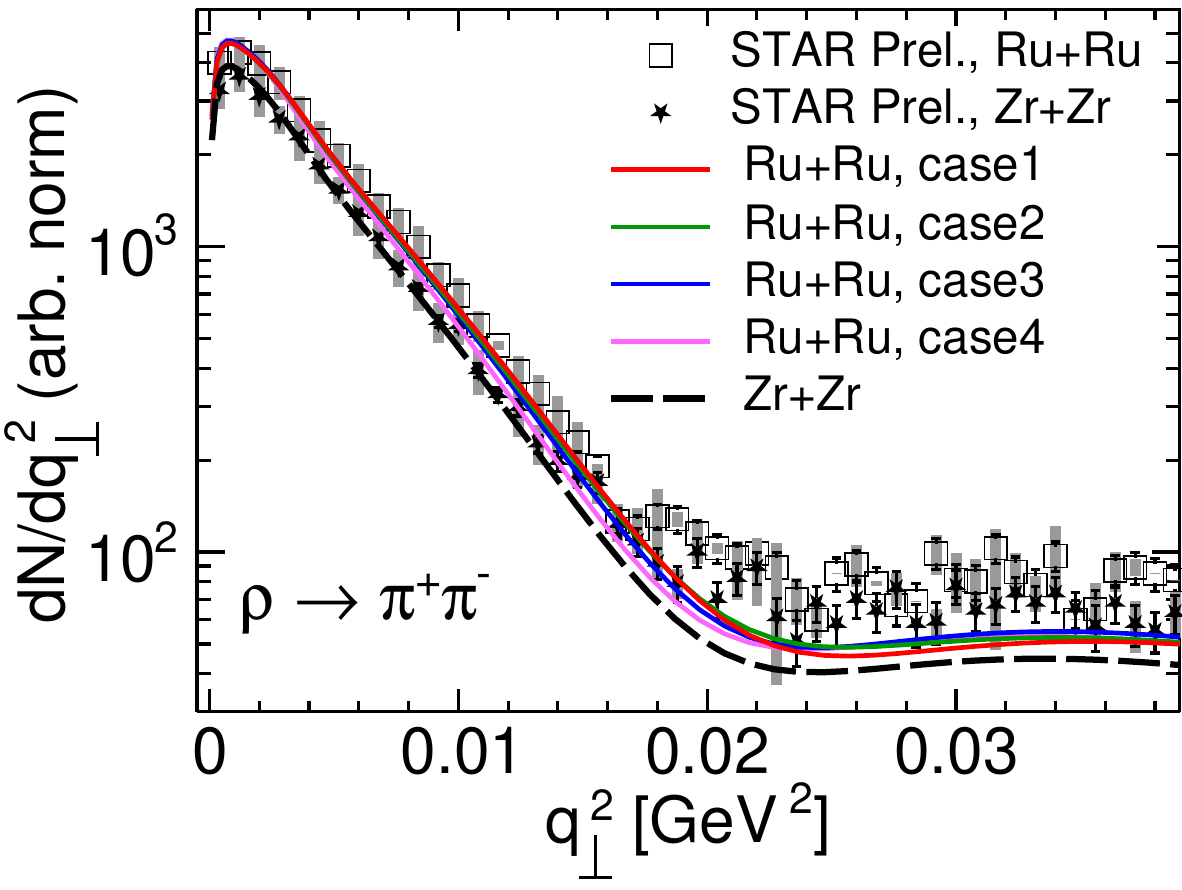}
    \caption{ Differential $\rho$ photoproduction cross-section in  ultra-peripheral Ru+Ru and Zr+Zr collisions using different parametrizations for the Ru geometry shown in Table~\ref{tab:isobartable}.
    The preliminary STAR data is from~\cite{STAR:2023zhao}}
\label{fig:isobarspectra}
\end{figure}

Fig.~\ref{fig:isobarspectra} shows the $q^2_{\perp}$-spectra of $\rho$ production in ultra-peripheral \RuRu\ and \ZrZr\ collisions, considering different deformation parameters for Ru and one set of parameters for Zr as shown in Table \ref{tab:isobartable}.  Following the STAR measurement \cite{STAR:2023zhao}, we include all possible Zero Degree Calorimeter (ZDC) neutron classes in the isobar calculations. With the default parameter sets (case 1 for Ru), our framework provides a reasonable description of the shape of the $q^2_{\perp}$-spectra for $\rho$ production. The agreement is particularly good for $q_\perp^2 < 0.015\,{\rm GeV}^2$, beyond which the data is underestimated for both collision systems. 

\begin{figure}
    \centering
    \includegraphics[width=0.44\textwidth]{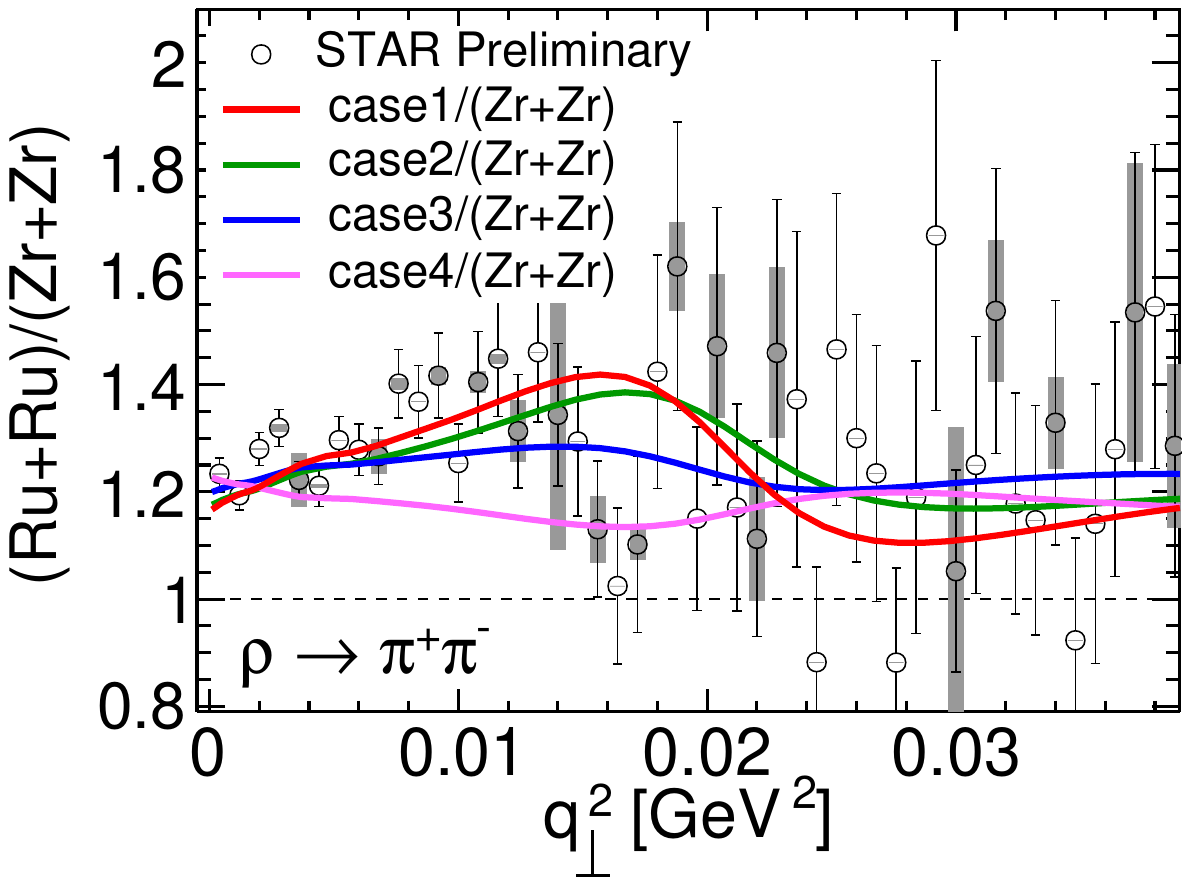}
    \caption{ The ratio of $\dd N/\dd q^2_{\perp}$ as a function of $q^2_{\perp}$ between Ru+Ru and Zr+Zr collisions with different parameter sets on Table.~\ref{tab:isobartable}. The preliminary STAR data is from~\cite{STAR:2023zhao}.}
\label{fig:isobarratio}
\end{figure}

Fig.~\ref{fig:isobarratio} presents the ratio between the $q_\perp^2$ spectra of the two systems, which helps to mitigate some systematic uncertainties, also in our calculation (e.g.~related to the vector meson wave function). Our results demonstrate that this ratio is sensitive to the specific choice of Woods-Saxon parameters. For instance, at $q^2_{\perp}\approx 0.016~{\rm GeV^2}$, the ratio reaches 1.44 for the (Ru+Ru, case 1)/(Zr+Zr) parameter choice, while it is 1.19 for case 4.  By comparing different cases, one can isolate the effects of the parameters $a_{\rm WS}$, $\beta_2$, and $\beta_3$, respectively. The difference between Ru+Ru, case 4, and Zr+Zr is primarily caused by the photon flux, which is proportional to the square of the atomic number ${\rm (Z_{Ru}/Z_{Zr})^2 = 1.21}$. 

The preliminary STAR data for the ratio exhibit a clear increasing trend for $q^2_{\perp} < 0.014~{\rm GeV^2}$, which aligns well with our calculation of the ratio between the case using the complete ruthenium parameter set and zirconium [(Ru+Ru, case 1)/(Zr+Zr)]. Cases 3 and 4 are clearly different from this result, with case 4 being the least compatible with the data. This indicates that the $\beta_2$ and diffuseness $a_{\rm WS}$ differences between isobars could be constrained using precise measurements of ratios like this.
At higher $q^2_\perp$ (where non-perturbatively large dipoles also have a non-negligible contribution) the different choices for Ru parameters result in very similar cross-section ratios and the STAR data can not distinguish between the different geometries.
These results emphasize the significance of considering the detailed nuclear structure in studying vector meson production in ultra-peripheral collisions of \RuRu\ and \ZrZr. The sensitivity to various deformation parameters opens up opportunities for further investigations and constraining the nuclear structure of isobar systems.

\section{Conclusions}
\label{sec:conclusions}
We have studied vector meson production in ultra-peripheral collisions at RHIC and the LHC involving Au+Au, U+U, Pb+Pb, and Ru--Zr isobar systems within the CGC framework. We investigated the $q^2_{\perp}$ dependence of the differential cross-sections and studied the $\cos(2\Delta\Phi)$ modulations in the momentum distribution of the decay particles. 

Including the interference effect and the transverse momentum of the linearly polarized photons in the UPC, our CGC calculations for vector meson production successfully reproduced the $q_\perp^2$ spectrum of $\rho$ production and the $\cos(2\Delta\Phi)$ azimuthal angular correlation of the decay pions in ultra-peripheral collisions both at RHIC and the LHC. 
For the angular modulation, it was essential to consider the linear polarization of the incoming photons.

We studied the dependence of the cross-section and its modulations on the nuclear species with a focus on the role of different radii and deformations. We identified that the sensitivity to variations of the nuclear radius and degree of deformation is mainly a result of different radii and deformations requiring different minimal impact parameters $B_{\rm min}$ between the two nuclei to obtain a UPC. Increasing $B_{\rm min}$ leads to weaker interference effects and smaller modulation amplitudes. We further reported the systematic uncertainty of our results from the lack of precise knowledge of $B_{\rm min}$.

We next extracted the nuclear radius as a function of the angle $\Delta\Phi$ by fitting the $q_\perp^2$ spectra. The extracted radius from the CGC calculations for Au nuclei is in good agreement with that obtained by the STAR Collaboration when using the default Woods-Saxon parameters for Au. 
For uranium nuclei, we need to assume a larger radius than the default value to describe the STAR data. When this is done, the extracted radius, which agrees with the STAR result, is larger than the U charge radius plus a neutron skin typical for similar nuclei.

Investigating the effects of nuclear deformation on the extracted radii, we found that more deformation leads to smaller extracted radii because i) more deformation leads to more fluctuations and thus an increased incoherent contribution, which makes the total spectra flatter; and ii) deformation alters the distribution of impact parameters, allowing in particular for smaller impact parameters that enter with a larger weight and go along with larger photon transverse momentum, which also flattens the spectra.

We further predicted the $\phi$ and \jpsi meson production cross-sections, along with the $\cos(2\Delta\Phi)$ modulations in their decay product distributions, for ultra-peripheral Au+Au collisions. The behavior of $\phi$ mesons resembled that of $\rho$ production due to the same scalar nature of their decay products and similar wave functions. In contrast, \jpsi mesons exhibited $\cos(2\Delta\Phi)$ modulations with an opposite sign to those of $\rho$ and $\phi$ mesons, which is attributed to the spin 1/2 nature of its decay products.

Compared to $\rho$ production data in Pb+Pb UPCs at LHC from the ALICE Collaboration, our calculations showed good agreement with the measured angular modulations. The trend with changing forward neutron multiplicity classes is well reproduced and again is dominated by the role of the minimally allowed impact parameter.

Using a ratio of cross-sections, we presented the sensitivity of the production of $\rho$ mesons in ultra-peripheral isobar collisions to the specific structures of Ru and Zr nuclei. Our calculation demonstrates the potential of this process as an innovative means to constrain the geometric features of isobaric nuclei.

Our study highlights the potential of vector meson production in ultra-peripheral collisions as a powerful tool for exploring spatial gluon distributions and geometric deformations of nuclei at small-$x$. 

\section*{Acknowledgments}
We thank  J. Zhao for providing the STAR data.
We are grateful to J. D. Brandenburg, A. G. Riffero,  A. I. Sheikh, Z. Xu, Z. Ye, W. Zha, C. Zhang, and Y.-J. Zhou for valuable discussions.
This material is based upon work supported by the U.S. Department of Energy, Office of Science, Office of Nuclear Physics, under DOE Contract No.~DE-SC0012704 (B.P.S.) and Award No.~DE-SC0021969 (C.S.), and within the framework of the Saturated Glue (SURGE) Topical Theory Collaboration.
C.S. acknowledges a DOE Office of Science Early Career Award. 
H.M. is supported by the Research Council of Finland, the Centre of Excellence in Quark Matter, and projects 338263 and 346567, and under the European Union’s Horizon 2020 research and innovation programme by the European Research Council (ERC, grant agreement No. ERC-2018-ADG-835105 YoctoLHC) and by the STRONG-2020 project (grant agreement No 824093) and wishes to thank the EIC Theory Institute at BNL for its hospitality during the completion of this work.
F.S.  is supported by the  National  Science  Foundation under grant  No. PHY-1945471, and partially supported by the UC Southern California Hub, with funding from the UC National  Laboratories division of the  University of  California  Office of the  President.
F.S. and W.B.Z. are supported by the National Science Foundation (NSF) under grant number ACI-2004571 within the framework of the XSCAPE project of the JETSCAPE collaboration.
W.B.Z is also supported by the US DOE under Contract No. DE-AC02-05CH11231, and within the framework of the SURGE Topical Theory Collaboration.
The content of this article does not reflect the official opinion of the European Union and responsibility for the information and views expressed therein lies entirely with the authors.
This research was done using resources provided by the Open Science Grid (OSG)~\cite{Pordes:2007zzb, Sfiligoi:2009cct}, which is supported by the National Science Foundation award \#2030508.

\bibliographystyle{JHEP-2modlong.bst}

\bibliography{spires}

\end{document}